\documentclass[aps,prb,twocolumn,nofootinbib,groupedaddress,superscriptaddress
,longbibliography]{revtex4-2}

\usepackage[utf8]{inputenc}
\usepackage{amsmath,amssymb}
\usepackage{graphicx}
\usepackage{multirow}
\usepackage{bm}
\usepackage{mathtools}
\usepackage{dsfont}
\usepackage{amsfonts}
\usepackage{xcolor}
\usepackage[normalem]{ulem}
\usepackage{url}
\usepackage{wasysym}
\usepackage{bbm}
\usepackage{hyperref}
\hypersetup{
    colorlinks=true,
    linkcolor=blue,
    filecolor=magenta,      
    urlcolor=blue,
    citecolor=blue,
}
\usepackage{braket}
\usepackage{booktabs}
\usepackage{float}

\newcommand{\ourtitle}{
Microscopic Fingerprint of Chiral Superconductivity}

\allowdisplaybreaks

\begin{document}
\title{\textbf{\ourtitle}}
\author{Xuefeng Wu} 
\altaffiliation{These authors contributed equally to this work}
\affiliation{School of Physical Sciences, Great Bay University, Dongguan 523000, China}

\author{Xuan Hao} 
\altaffiliation{These authors contributed equally to this work}
\affiliation{School of Physical Sciences, Great Bay University, Dongguan 523000, China}
\affiliation{State Key Laboratory of Quantum Functional Materials, Department of Physics, Southern University of Science and Technology, Shenzhen 518055, China}

\author{Zhuo Chen} 
\altaffiliation{These authors contributed equally to this work}
\affiliation{Department of Physics and Astronomy, The University of Tennessee, Knoxville, Tennessee 37996, USA}

\author{Yuchang Cai} 
\affiliation{Department of Physics and Astronomy, The University of Tennessee, Knoxville, Tennessee 37996, USA}

\author{Minghao Wu} 
\affiliation{State Key Laboratory of Quantum Functional Materials, Department of Physics, Southern University of Science and Technology, Shenzhen 518055, China}

\author{Congrun Chen} 
\affiliation{State Key Laboratory of Quantum Functional Materials, Department of Physics, Southern University of Science and Technology, Shenzhen 518055, China}

\author{Kedong Wang} 
\email{wangkd@sustech.edu.cn}
\affiliation{State Key Laboratory of Quantum Functional Materials, Department of Physics, Southern University of Science and Technology, Shenzhen 518055, China}

\author{Fangfei Ming} 
\email{mingff@mail.sysu.edu.cn}
\affiliation{State Key Laboratory of Optoelectronic Materials and Technologies, School of Electronics and Information Technology and Guangdong Province Key Laboratory of Display Material, Sun Yat-sen University, Guangzhou, China}

\author{Steven Johnston}
\affiliation{Department of Physics and Astronomy, The University of Tennessee, Knoxville, Tennessee 37996, USA}
\affiliation{Institute for Advanced Materials and Manufacturing, The University of Tennessee, Knoxville, Tennessee 37996, USA}

\author{Rui-Xing Zhang}
\email{ruixing@utk.edu}
\affiliation{Department of Physics and Astronomy, The University of Tennessee, Knoxville, Tennessee 37996, USA}
\affiliation{Institute for Advanced Materials and Manufacturing, The University of Tennessee, Knoxville, Tennessee 37996, USA}
\affiliation{Department of Materials Science and Engineering, The University of Tennessee, Knoxville, Tennessee 37996, USA}

\author{Hanno H. Weitering}
\email{hanno@utk.edu}
\affiliation{Department of Physics and Astronomy, The University of Tennessee, Knoxville, Tennessee 37996, USA}
\affiliation{Institute for Advanced Materials and Manufacturing, The University of Tennessee, Knoxville, Tennessee 37996, USA}

\begin{abstract}
Chiral superconductors have long been theorized to break time-reversal symmetry and support exotic topological features such as Majorana modes and spontaneous edge currents, promising ingredients for quantum technologies. Although several unconventional superconductors may exhibit time-reversal symmetry breaking, clear microscopic evidence of chiral pairing has remained out of reach. In this work, we demonstrate direct real-space signatures of chiral superconductivity in a single atomic layer of tin on Si(111). Using quasiparticle interference imaging, we detected symmetry-locked nodal and antinodal points in the Bogoliubov quasiparticle wavefunction, tightly bound to atomic point defects in the tin lattice. These nodal features, along with their surrounding texture, form a distinct real-space pattern exhibiting a clear and exclusive hallmark of chiral superconductivity. Our findings, reinforced by analytical theory and numerical simulations, offer unambiguous evidence of chiral pairing in a two-dimensional material.
\end{abstract}

\maketitle

\let\oldaddcontentsline\addcontentsline
\renewcommand{\addcontentsline}[3]{}

Chiral superconductivity is a long-sought phase of matter in which Cooper pairs acquire a well-defined handedness, spontaneously breaking time-reversal symmetry. These states are predicted to host Majorana bound states, protected edge modes, and exotic vortices—hallmarks of topological superconductivity with potential applications in quantum information processing~\cite{1ReadPaired2000,2SankarNon-Abelian2008,3satotopological2017}. Despite extensive efforts, definitive microscopic evidence for chiral pairing has remained elusive. Leading candidate materials, including $\rm Sr_2RuO_4$~\cite{4luketime-reversal1998,5MackenzieThesuperconductivity2003,6XiaHigh2006,7pustogowconstraints2019,8AaronEvidence2021}, $\rm UTe_2$~\cite{9aokiunconventional2019,10ShengNearly2019,11QiangqiangPair2025}, and a range of heavy-fermions~\cite{12ERSchemmObservation2014,13aversbroken2020} and engineered platforms~\cite{14ARibakChiral2020,15SYFrankZhaoTime-reversal2023,16wanunconventional2024,17hansignatures2025}, exhibit signatures of time-reversal symmetry breaking but lack unambiguous confirmation of chiral phase winding.

\begin{figure}[htbp]
    \includegraphics[width=0.5\columnwidth]{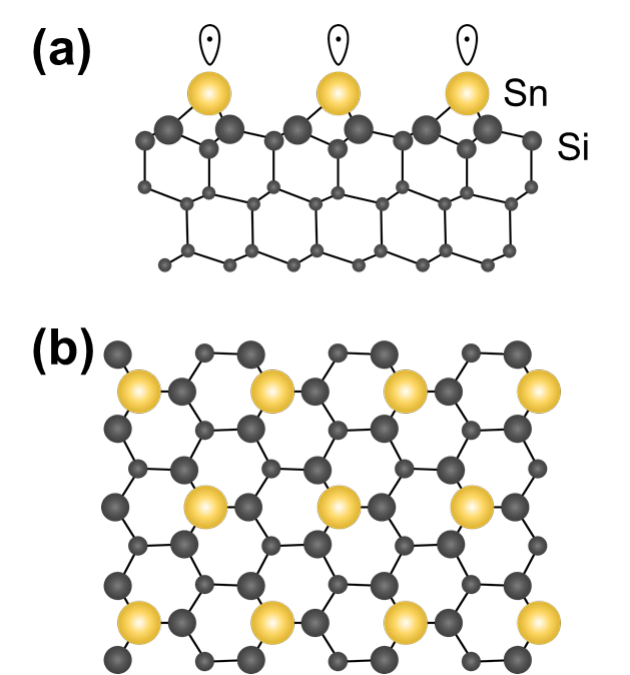}
    \caption{(a) Top and (b) side views of the Sn/Si(111) interface. The Sn atoms (yellow) adsorb right above the second-layer Si atoms (small gray spheres). The Sn-Sn adatom spacing is $\sqrt{3}\times3.84~\rm{\mathring{A}}=6.65~\rm{\mathring{A}}$ where $3.84~\rm{\mathring{A}}$ is the Si-Si distance in the Si(111) plane. } 
    \label{fig1}
\end{figure}

\begin{figure*}[t] 
    \centering
    \includegraphics[width=0.9\textwidth]{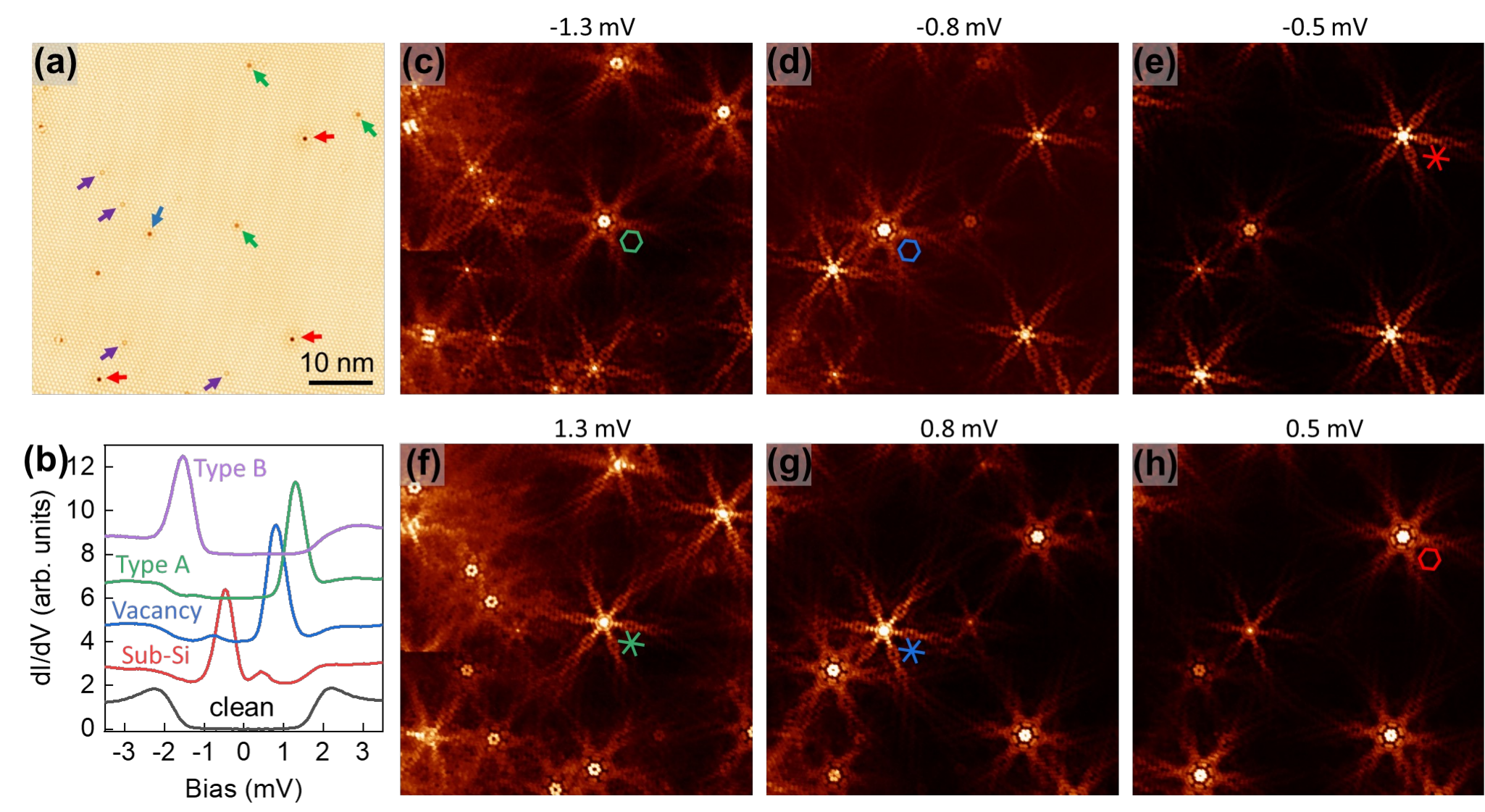}
    \caption{(a) Atomically resolved $55\times55 ~\rm{nm}^2$ STM image of Sn adatoms on the Si(111) surface (sample bias $V_s=20~\rm meV$; tunneling current $I_t=0.1~\rm nA$). Blue arrows indicate atomic-scale vacancies in the Sn lattice, revealing the underlying Si atoms. Red arrows mark substitutional defects (`Sub-Si') where a Sn adatom is replaced by a Si atom. Green and purple arrows highlight unidentified `Type A' and `Type B' point defects, respectively (see Fig.~\ref{Extendfig2}). (b) $dI/dV$ spectra acquired at the defect sites indicated in panel (a). Their color coding matches the arrow colors in panel (a). Each spectrum exhibits a pair of in-gap bound states straddling the Fermi level, with pronounced asymmetry in their intensities. A reference spectrum from a defect free region is included at the bottom, displaying a well-defined superconducting gap. (c)-(h) $dI/dV$ maps of same surface region, recorded at the energies of the bound states identified in panel (b). The maps reveal star-like and hexagon-like spatial features at opposite bias polarities, with the sequence of these patterns depending on the specific bound state energy. Visual markers next to point defects are added to highlight common LDOS features.
    }
    \label{fig2}
\end{figure*}

\begin{figure*}[t] 
    \centering
    \includegraphics[width=0.9\textwidth]{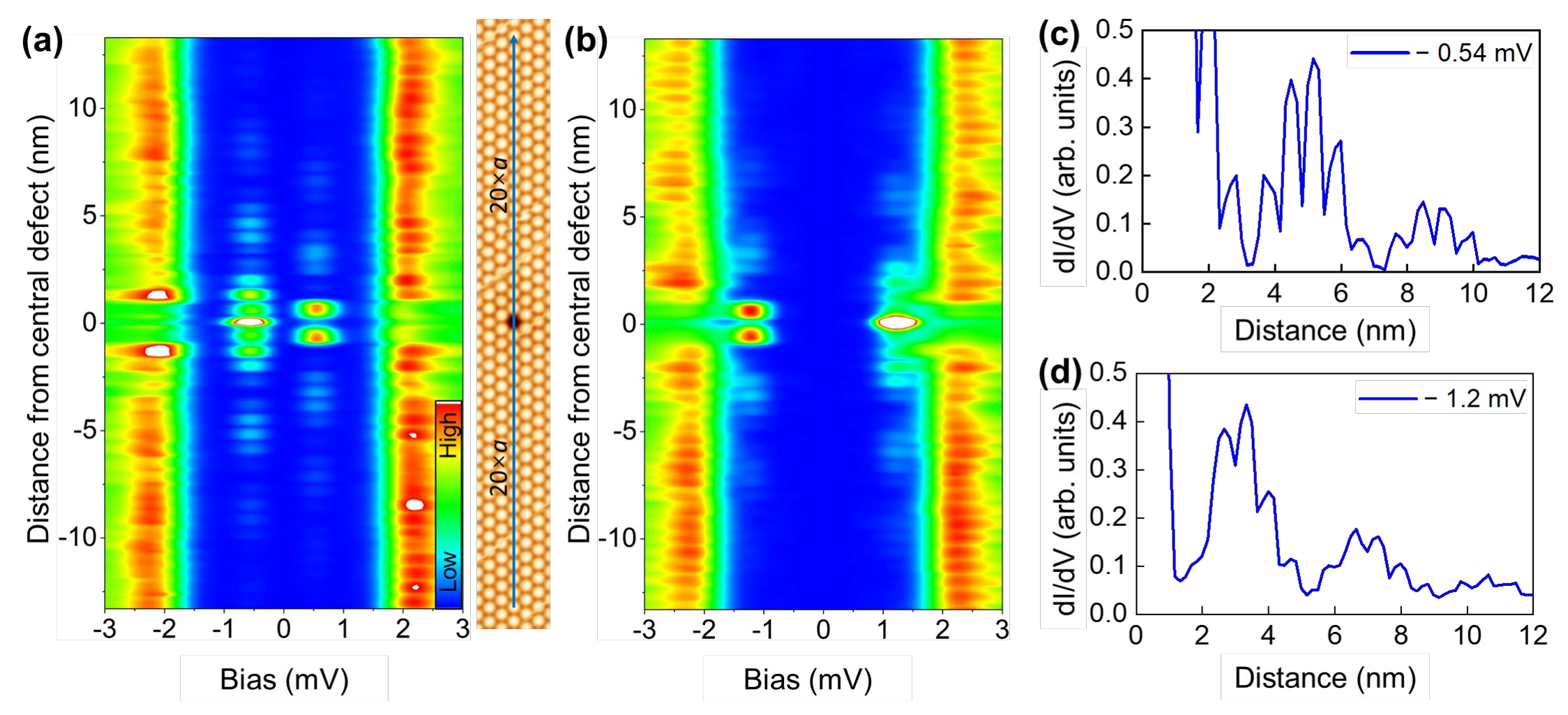}
    \caption{$dI/dV$ intensity mapping of in-gap bound states associated with (a) a substitutional ($\pm0.54~\rm meV$) and (b) `Type A' point defect ($\pm1.20~\rm meV$). The intensities are extracted from a series of single-point $dI/dV$ spectra measured along the row of Sn atoms indicated in panel (a) with the point defect at the center. The broad blue band represents the superconducting gap. Pronounced peaks (red) at $\pm2.16~\rm meV$ correspond to the coherence peaks. The intensities of the in-gap bound states decay with distance from the central defect site, exhibiting a superimposed oscillatory beating pattern with wavevectors $\boldsymbol{q}_1=2\boldsymbol{k}_\text{F}$ and $\boldsymbol{q}_2=\boldsymbol{G}_{11}-2\boldsymbol{k}_\text{F}$ (see Fig.~\ref{Extendfig3}; $\boldsymbol{G}_{11}$ is a reciprocal lattice vector). The oscillations at positive and negative bias are largely out of phase. (c) and (d), Corresponding $dI/dV$ intensities plotted as a function of distance from the defect, revealing two characteristic spatial modulations with approximate periodicities of $7~\rm{\mathring{A}}$ and $34~\rm{\mathring{A}}$.
    }
    \label{fig3}
\end{figure*}

Quasiparticle interference (QPI) imaging using scanning tunneling microscopy (STM) is a powerful technique for probing superconducting pairing symmetries. In systems such as $\rm NbSe_2$~\cite{18AnjanQuantum2013}, Fe(Se,Te)~\cite{19ZhenyuEvidence2020,20THanaguriUnconventional2010}, and the cuprate superconductors~\cite{21KMcElroyAtomic-Scale2005}, QPI has provided critical insights into the momentum dependence of the superconducting gap, coherence factors, and the sign structure of the order parameter. However, interpreting QPI patterns remains challenging, especially in multiband systems where analysis often requires input from angle-resolved photoemission spectroscopy (ARPES) data and theoretical modeling~\cite{22allanimaging2013,23GaoPossible2018,24bokerquasiparticle2019}. Further complicating interpretation is the often-unknown nature and range of the scattering centers and tunneling pathways~\cite{25KreiselInterpretation2015}, which can obscure the connection between observed QPI features and the underlying pairing symmetry. 

The Sn/Si(111) platform offers a key advantage in its structural and electronic simplicity (Fig.~\ref{fig1}). It consists of a triangular array of Sn atoms adsorbed on the Si(111) surface. Each Sn atom forms three backbonds with the Si substrate, leaving one half-filled dangling bond orbital per Sn site. This dangling-bond-derived surface state lies entirely within the Si band gap~\cite{31MingRealization2017}. The surface state bandwidth $W\cong0.5~\rm eV$ and Hubbard interaction energy $U\cong0.66~\rm eV$ indicate a Mott insulating state for the undoped system~\cite{30limagnetic2013,31MingRealization2017}. In hole-doped samples, superconductivity emerges from a Cooper instability in the dangling bond orbitals of Sn adatoms, with a critical temperature $T_c$ of 9~K at $\sim$6\% doping~\cite{26WuXuefengSuperconductivity2020,27mingevidence2023}. As a doped single-band Mott insulator~\cite{28ProfetaTriangular2007,29ModestiInsulating2007,30limagnetic2013,31MingRealization2017,32JageralphaSn2018}, the system exhibits cuprate-like behavior, where strong correlations are believed to drive $d$-wave pairing~\cite{33ScalapinoAcommon2012}. However, the triangular symmetry of the Sn adatom lattice supports a chiral variant, $d_{x^2-y^2}\pm id_{xy}$ that breaks time-reversal symmetry~\cite{27mingevidence2023,34SigristPhenomenological1991,35kallinchiral2016,36Zamproniochiral2023,37CaoXiaodongChiral2018}. Prior momentum-space QPI studies have pointed to such symmetry breaking~\cite{27mingevidence2023}, while enhanced zero-bias conductance at domain boundaries suggests the presence of edge states. However, their topological character could not be verified and the definitive proof of chirality remained an open question~\cite{27mingevidence2023}. 

Because STM provides real-space access to the relevant pairing orbitals, including those at atomic defects, high-resolution QPI imaging should, in principle, unveil signatures of the underlying pairing symmetry—provided that the scattering potential and tunneling pathway are sufficiently well understood and modeled. Here, we identify a microscopic fingerprint of chiral Cooper pairing from high-quality real-space QPI data centered at atomic-scale point defects in the Sn layer. Remarkably, the scattering potential at these defects is extremely simple—well described by a delta function—while electrons tunnel directly into the active layer. The observed anisotropic interference patterns are almost perfectly reproduced without parameter fine-tuning. As we will show, this level of agreement between our modeling and the experiments gives us unprecedented ability to confirm a chiral order parameter and rule out any non-chiral pairing variants.

\begin{figure}[t]
    \includegraphics[width=1.0\columnwidth]{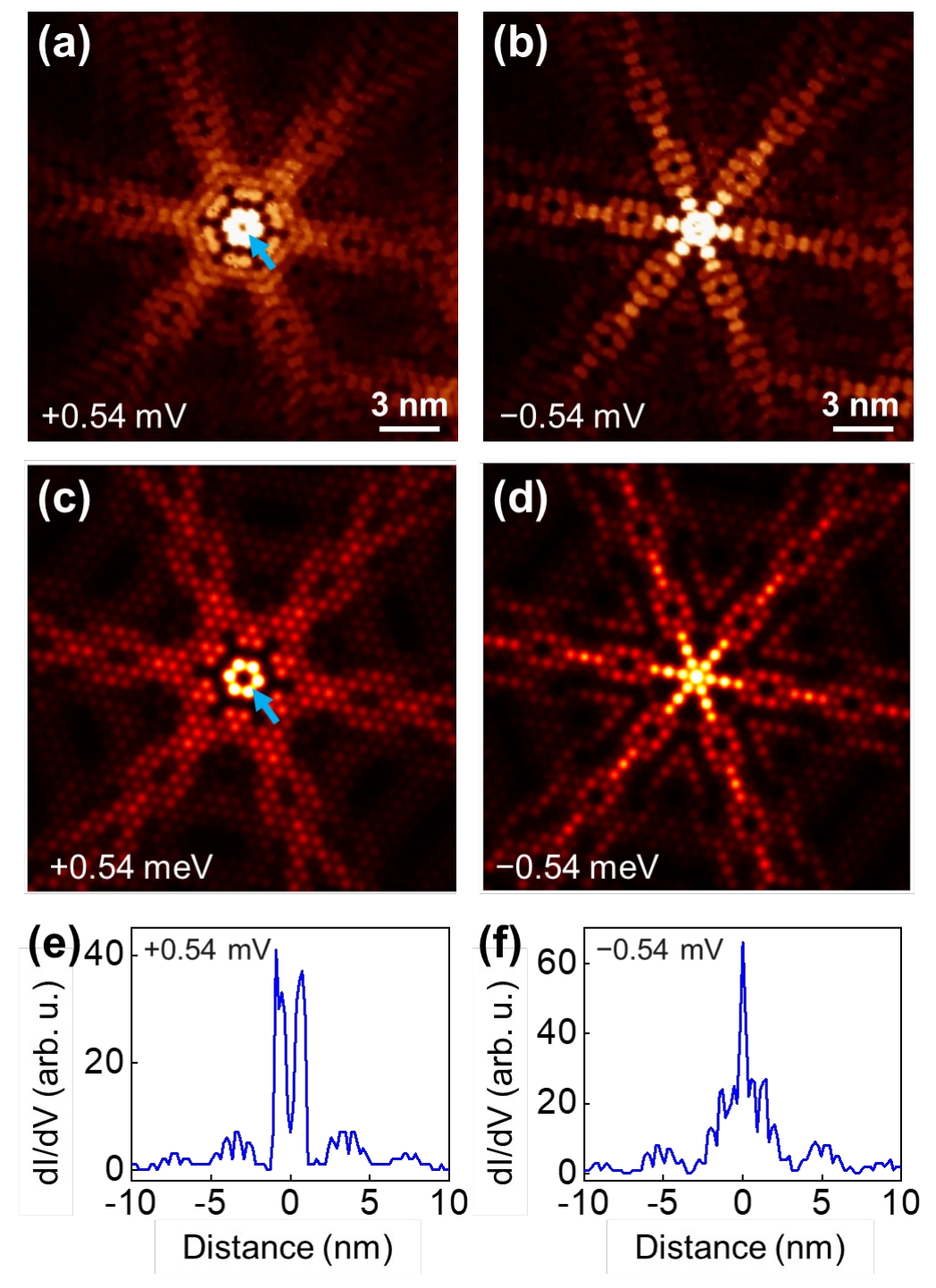}
    \caption{(a) and (b) Experimental $dI/dV$ maps of a substitutional Si defect, located at the center of each image. (c) and (d) Theoretical LDOS maps calculated for a repulsive scattering potential of $183~\rm meV$ (see Fig.~\ref{fig5}). Notice the pronounced dark spot at the center of panels (a) and (c), indicated by blue arrows, which is a key signature of a chiral superconducting order parameter. (e) and (f) $dI/dV$ line scans along the atom rows, showing the strong node or antinode at the defect site. The remaining small $dI/dV$ signal at the antinode location is attributed to the finite spatial resolution of STM.} 
    \label{fig4}
\end{figure}

Fig.~\ref{fig2}(a) shows an atomically resolved $55\times55~ \rm{nm}^2$ topographic image of the Sn/Si(111) system, acquired at 400~mK using a tungsten tip. The image reveals a highly ordered Sn adatom lattice interspersed with randomly distributed atomic-scale point defects, marked by arrows. Scanning tunneling spectroscopy ($dI/dV$ point spectra) of the unperturbed Sn sites exhibits a `hard' U-shaped superconducting gap with sharp coherence peaks at –2.26~meV and +2.22~meV (Fig.~\ref{fig2}(b), bottom spectrum). This gap is well described by either a chiral $d$-wave or an anisotropic $s$-wave order parameter. In contrast, a conventional isotropic $s$-wave model yields a poor fit, while single-component $d$-wave and chiral $p$-wave symmetries can be excluded due to their characteristic soft energy gaps (see Appendix B).

Each point defect generates a pair of in-gap bound states, confirming their pair-breaking nature (Fig.~\ref{fig2}b). The defects can be classified by their bound-state energy. Substitutional Si defects (red) exhibit bound states around $\pm0.54~\rm meV$, while vacancy defects (blue) show states around $\pm0.85~\rm meV$. In addition, there are two unknown point defects, labeled `Type A' (green) and `Type B' (purple) defects (see Appendix C for defect identification). $dI/dV$ line scans taken along the atom rows (Fig.~\ref{fig3}) reveal that these bound states extend up to $\sim$12~nm from the defect center. The intensity of the $dI/dV$ signal exhibits an oscillatory beating pattern governed by two fundamental wavevectors, $\boldsymbol{q}_1=2\boldsymbol{k}_\text{F}$ and $\boldsymbol{q}_2=\boldsymbol{G}_{11}-2\boldsymbol{k}_\text{F}$, where $|\boldsymbol{k}_\text{F}|=0.38\pm0.01~\rm{\mathring{A}}^{-1}$ is the Fermi wave vector measured along the $\overline{\Gamma K}$ direction in reciprocal space and $\boldsymbol{G}_{hk}$ is a reciprocal lattice vector. These wavevectors are clearly visible in the momentum-space QPI maps (Appendix D), connecting nearly parallel segments of the Fermi surface and its replicas. The exceptionally long-range Friedel oscillations observed here are reminiscent of magnetic-impurity bound states in conventional $s$-wave superconductors such as Pb~\cite{38RubyOrbital2016}, 2H-NbSe$_2$~\cite{39menardcoherent2015}, and La(0001)~\cite{40kimlong-range2020}, which also exhibit similar Fermi surface topologies—though without the distinct beating patterns reported here.

\begin{figure*}[t] 
    \centering
    \includegraphics[width=\textwidth]{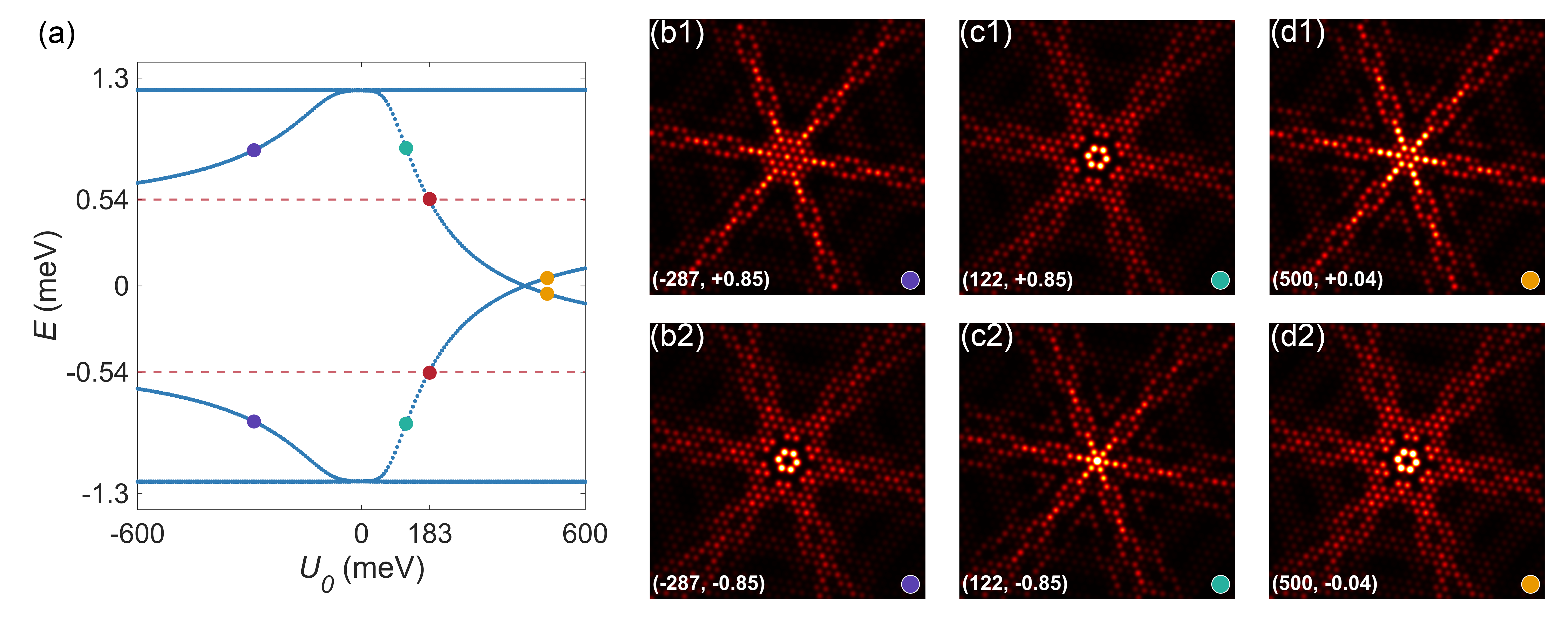}
    \caption{(a) Bound state energies $E_0$ of point defects as a function of impurity potential strength, calculated using Bogoliubov–de Gennes (BdG) theory for a chiral $d$-wave order parameter. For attractive potentials ($U_0<0$), the bound state energies asymptotically approach the center of the superconducting gap with increasing potential strength. In contrast, for repulsive potentials, the bound states cross the Fermi level at a critical potential strength of $438~\rm meV$. The LDOS maps of the substitutional defect shown in Fig.~\ref{fig4} were computed using $U_0 = 183~\rm meV$ (red dots in panel (a)), chosen to reproduce the experimentally observed bound state energies at $\pm0.54~\rm meV$. (b–d) Additional LDOS simulations for impurity potentials indicated by the colored dots in a, labeled by $(U_0, E_0)$. In particular, an attractive impurity with $U_0=-287~\rm meV$ (purple dots) yields bound states at $\pm0.85~\rm meV$, consistent with the natural vacancy defects in Fig.~\ref{fig2}. Notably, both the node-antinode and flower-star spatial patterns persist across a broad range of impurity strengths, consistent with experimental observations. Their polarity inverts both with the sign change of the scattering potential and upon crossing the quantum critical point at 438 meV. Notably, for $U_0=-287~\rm meV$, the star-like LDOS (b1) appears more isotropic within the second hexagonal shell surrounding the impurity, with the six arms becoming noticeably blurred, agreeing with the experimental LDOS in Fig.~\ref{fig2}(g). In contrast, both simulation and experimental LDOS maps of substitutional Si defects produce a sharper, more anisotropic star pattern. This atomic-level agreement of these LDOS simulations provides further evidence for intrinsic chiral $d$-wave superconductivity in Sn/Si(111).}
    \label{fig5}
\end{figure*}

The real-space QPI pattern reflects the spatial distribution of the local density of states (LDOS $\propto$ $dI/dV$). The LDOS maps in Fig.~\ref{fig2}(c–h), acquired over the same area shown in Fig.~\ref{fig2}(a), reveal the structure of defect-bound states with significantly improved spatial and spectroscopic resolution compared to our previous report~\cite{27mingevidence2023}. As shown in Fig.~\ref{fig2}(b), the in-gap bound states are symmetrically located about the Fermi level, at $E_\text{F}\pm E_0$. One state displays a sharp, star-like pattern with spikes aligned along the atom rows (or $\overline{\Gamma K}$ direction), reflecting long-range Friedel oscillations. The other, at opposite polarity, exhibits a more rounded, hexagonal, flower-like structure. These features are indicated by visual markers next to representative defects.

Remarkably, the spatial features are identical across all three observed point defects, despite large variations in bound-state energies. This consistency indicates that the LDOS maps reflect intrinsic properties of the Sn/Si(111) system rather than the specific nature of each defect. Notably, the star-like and flower-like patterns can reverse polarity depending on the type of defect. The corresponding bound-state energies should depend on the strength of the scattering potential~\cite{41WangImpurity2004,42mashkooriimpurity2017}. While $\boldsymbol{q}$-space QPI images primarily emphasize the pole structure of the electronic response function, governed by the Fermi surface topology, the real-space LDOS maps provide a direct and distinctive fingerprint of the superconducting order parameter.

Fig~\ref{fig4}(a-b) shows high-resolution real-space images of bound states associated with a substitutional Si defect. At the positive bound-state energy $E_0=0.54~\rm meV$, the LDOS shows an atomically sharp node at the impurity site, indicated by the arrow in panel (a), surrounded by a bright hexagonal ring formed by enhanced LDOS on the six nearest neighbors. A dip at the next-nearest neighbors follows, producing a ripple pattern that evolves into long-range Friedel oscillations (Fig.~\ref{fig3}). In contrast, the corresponding state at $-E_0$ in panel (b) exhibits a more centralized, star-like pattern with LDOS maxima at the defect site and adjacent atoms. In fact, the large anti-nodal peak at the defect site (Fig.~\ref{fig4}b) sharply contrasts with the nodal behavior of the LDOS at $+E_0$ in Fig.~\ref{fig4}a. This observation is perhaps even more striking in the $dI/dV$ line scan in Figs.~\ref{fig4}e and \ref{fig4}f. This characteristic defect-center node-antinode and “flower-star” dichotomy persists across defect types and bound state energies and is therefore robust.

To understand this striking pattern, we perform large-scale real-space simulations with an {\it ab-initio}-based tight-binding model of Sn/Si(111)~\cite{43MarchettiElectronic2025}. Details of the model Hamiltonian and simulation procedure are provided in the Methods section. Assuming chiral $d$-wave pairing order, we introduce a delta-function-like nonmagnetic impurity potential with strength $U_0=183~\rm meV$, chosen to reproduce the experimentally observed bound-state energies of the substitutional Si defect, as shown in Fig.~\ref{fig5}. Remarkably, the simulated LDOS patterns in Figs.~\ref{fig4}(c–d) convincingly capture all key experimental features, including the contrasting node-antinode behavior and flower-star textures. Importantly, this agreement remains robust over a wide range of $U_0$, in line with the consistency of the experimental results across various impurities (Fig.~\ref{fig5}). The high level of correspondence—achieved without fine-tuning—underscores the predictive power of our modeling framework, in which all parameters are derived from first-principles calculations or experimental data.

The reproduction of the LDOS patterns naturally invites a closer look at their microscopic origin. In a recent complementary study~\cite{44caiDeciphering2025}, we demonstrate that such robust, particle–hole asymmetric nodal behavior of LDOS is a generic hallmark of chiral superconductors with atomically localized impurities. Specifically, for chiral $d\pm id$ pairing on a $C_6$-symmetric Sn lattice, the pairing order carries an angular momentum of $\pm2$ and transforms nontrivially under rotations. As a result, the anomalous Green’s function, which encodes pairing correlations, vanishes exactly at the impurity site, as proved in the Methods section (Appendix A). This enforces a local decoupling of particle and hole sectors of the Bogoliubov bound states, forcing the impurity state to host either an electron component $\psi_e$ or a hole component $\psi_h$ at the impurity site, but not both. Since $\psi_e$ and $\psi_h$ give rise to LDOS resonances at $-E_0$ and $+E_0$, respectively, there will always exist an LDOS node at one energy and an antinode at the opposite energy, a direct consequence of the chiral Cooper pairs.

To rule out competing pairing scenarios, we repeat the simulations for conventional and (anisotropic) $s$-wave as well as chiral $p$-wave order parameters (Appendix E). In the $s$-wave case, only magnetic scatterers generate in-gap bound states; we therefore consider many possible combinations of magnetic and nonmagnetic impurity potentials. At first glance, all resulting images display the characteristic star-like features arising from the Fermi surface topology. However, a closer inspection reveals that the impurity-induced LDOS qualitatively differs from the experimental observations—the star-like patterns are all different and none of the simulated patterns reproduces the distinct flower–star motif seen in the data. In particular, the flower-like pattern of Fig.~\ref{fig4}a appears unique to chiral $d$-wave pairing. Non-chiral pairing states, in general, lack a robust (non-accidental) node–antinode structure at the impurity site. These discrepancies underscore the incompatibility of alternative pairing symmetries with the observed impurity response. Taken together, the experimental results, microscopic theory, and comparative simulations provide strong evidence that the superconducting state in Sn/Si(111) is characterized by intrinsic chiral $d$-wave pairing.

The definitive identification of chiral superconductivity presented here is made possible by the structural simplicity of the Sn/Si(111) system, which permits unambiguous detection of atomic-scale point defects that preserve the underlying $C_6$ symmetry and greatly facilitates modeling of the corresponding scattering potentials. In contrast, more complex or disordered defect configurations—such as an intentionally created vacancy via STM manipulation (Appendix F)—tend to obliterate the characteristic chiral LDOS signature due to symmetry breaking and the increased complexity of the scattering potential. This highlights a broader challenge: the chiral QPI fingerprint may remain hidden in many material systems. Our emphasis on symmetry-preserving point defects not only enables direct visualization of the chiral fingerprint in this work, but may also serve as a guiding principle for future QPI-based studies of unconventional superconductivity. Most importantly, the clear manifestation of chiral QPI patterns here provides unambiguous experimental evidence that topological superconductivity is a physically realizable quantum phase. Our findings also lend strong credence to recent theoretical predictions not only of chiral $d$-wave pairing at moderate hole doping, but also of proximate chiral $p$-wave and $f$-wave instabilities~\cite{45WolfTriplet2022,46BiderangTopological2022,47KimInterplay2023}, suggesting that Sn/Si(111) may serve as a versatile platform for exploring a rich landscape of exotic superconducting states.

\begin{acknowledgments}
The authors thank Elbio Dagotto for insightful discussions. K.W. acknowledges financial support from the Guangdong Basic and Applied Basic Research Foundation (Grant No. 2024A1515013040). X.W. and F.M. acknowledge financial support from the National Natural Science Foundation of China under Grant Nos. 12204225 and 12174456, respectively. Z.C., S.J., R.-X.Z. and H.H.W. were supported by the National Science Foundation Materials Research Science and Engineering Center (MRSEC) program through the UT Knoxville Center for Advanced Materials and Manufacturing under Grant No. DMR-2309083.
\end{acknowledgments}

\section*{Author Contributions}
XW, XH, MW, and CC carried out the experimental measurements. ZC and YC performed the numerical and analytical calculations, respectively. KW and FM co-supervised the experiment. RXZ led the theory project. HHW and RXZ wrote the manuscript with input from FM and SJ. HHW is the overall project lead.

\section*{Data availability}
The data supporting this study are available via Zenodo at https://doi.org/10.5281/zenodo.15741993. The code supporting this study will be deposited in a public repository before the paper is published online. Before that time, code will be made available upon request to the corresponding authors.

\appendix

\section{Methods}

\subsection{Sample preparation}
The sample preparation procedure follows established protocols from previous studies~\cite{26WuXuefengSuperconductivity2020,27mingevidence2023}. Briefly, a hole-doped Sn adatom layer was grown on a boron-doped Si(111) substrate with a nominal room-temperature resistivity of $0.005~\Omega\cdot \rm cm$. Hole doping is achieved via charge transfer between the surface layer and the bulk substrate, resulting in an estimated carrier concentration of approximately 0.08 holes per Sn atom~\cite{27mingevidence2023}. The Si substrate was annealed to $1200^{\circ}\rm C$ in ultrahigh vacuum to obtain an atomically clean surface. Sn atoms were deposited onto the Si(111) surface from a thermal effusion cell while maintaining the substrate temperature at approximately $600^{\circ}\rm C$. This process led to the formation of coexisting superconducting $(\sqrt{3}\times\sqrt{3})$ and semiconducting $(2\sqrt{3}\times2\sqrt{3})$ domains within the Sn adatom layer. The superconducting domains reached lateral dimensions of up to $300\times300 ~\rm{nm}^2$.

\subsection{Scanning Tunneling Microscopy and Spectroscopy measurements}

STM measurements were performed using a cryogenic STM system (Unisoku) capable of cooling both the sample and the tip to approximately 400~mK, under magnetic fields up to 15~T. Differential conductance spectra, including single-point $dI/dV$ point measurements and two-dimensional conductance maps, $g(\boldsymbol{r},V)\propto\rm LDOS$, were acquired using standard lock-in detection with a typical modulation voltage of $V_\text{rms}=0.14~\rm meV$ and a modulation frequency of 673~Hz. A typical $g(\boldsymbol{r},V)$ map consists of $240\times240$ pixels, covering a $55\times55 ~\rm{nm}^2$ area. The $\boldsymbol{q}$-space QPI images were obtained by computing the power spectral density of the Fourier transformed conductance maps $|g(\boldsymbol{r},V)|$.

\subsection{Impurity state simulations}

We model the electronic structure of Sn with a two-dimensional tight-binding model on a triangular lattice,
\begin{eqnarray}
\mathcal{H} = \sum_{i,j,s} t_{ij} c^{\dagger}_{i,s} c^{\phantom\dagger}_{j,s} + \sum_{i,j,s,s'} i \alpha_{ij} \left[\hat{\boldsymbol{z}}\cdot \boldsymbol{\sigma} \times (\boldsymbol{r}_i - \boldsymbol{r}_j) \right]c^{\dagger}_{i,s} c^{\phantom\dagger}_{j,s'}, \nonumber \\
\end{eqnarray}
where $c_{i,s}^{\dagger}$ creates an electron at lattice site $r_i$ with a spin index of $s=\uparrow,\downarrow$. Besides the spin-conserving electron hopping process, we also include the Rashba spin-orbital coupling effect $\alpha_{ij}$ with the spin Pauli matrices denoted as $\boldsymbol{\sigma}=(\sigma_x,\sigma_y,\sigma_z)$. We consider the nearest, next nearest, and third nearest neighboring hopping, denoted by $t_1=52.773~\rm meV$, $t_2=-0.2703t_1$, $t_3=0.0974t_1$, respectively~\cite{43MarchettiElectronic2025}. Together with a nearest-neighboring $\alpha_1=0.1029t_1$ and a chemical potential of $\mu=-22~\rm meV$, we find that the above parameter set quantitatively reproduces the Fermi surface revealed in the $\boldsymbol{k}$-space QPI measurement~\cite{27mingevidence2023}. Upon a Fourier transform, we have $\mathcal{H}=\sum_{\boldsymbol{k}}\psi_{\boldsymbol{k}}^\dagger h_0(\boldsymbol{k})\psi^{\phantom\dagger}_{\boldsymbol{k}}$ with $\psi_{\boldsymbol{k}}=(c_{\boldsymbol{k},\uparrow},c_{\boldsymbol{k},\downarrow} )^T$. To describe superconductivity, we further update the Hamiltonian matrix $h_0(\boldsymbol{k})$ to a Bogoliubov-de Gennes (BdG) form, with
\begin{equation}
\label{eq2}
H(\boldsymbol{k}) = 
\begin{bmatrix}
h_0(\boldsymbol{k}) & \Delta(\boldsymbol{k}) \\
\Delta^\dagger(\boldsymbol{k}) & -h_0^T(-\boldsymbol{k})
\end{bmatrix}.
\end{equation}

For chiral $d$-wave pairing, the pairing matrix $\Delta(\boldsymbol{k})=\Delta_d[f_{d,x}(\boldsymbol{k}) \sigma_x+f_{d,y}(\boldsymbol{k})\sigma_y]$, where the form factors $f_{d,x}(\boldsymbol{k})=0$ and $f_{d,y}(\boldsymbol{k}) = 2i[\cos(k_x) - \cos\left(\sqrt{3}k_y/2\right)\cos\left(k_x/2\right)] - 2\sqrt{3} \sin\left(\sqrt{3}k_y/2\right) \sin\left(k_x/2\right)$. For our purpose, we also consider two other candidate pairing channels: (1) $s$-wave with $\Delta(\boldsymbol{k})=i\Delta_s\sigma_y$ and (2) chiral $p$-wave with $\Delta(\boldsymbol{k})=\Delta_p [f_{p,x}(\boldsymbol{k}) \sigma_x+f_{p,y}(\boldsymbol{k})\sigma_y]$ and $f_{p,x}(\boldsymbol{k}) = 2\sqrt{3} \sin\left(\sqrt{3}k_y/2\right) \cos\left(k_x/2\right) + 2i \left[\sin(k_x) + \cos\left(\sqrt{3}k_y/2\right) \sin\left(k_x/2\right) \right],$ and $f_{p,y}(\boldsymbol{k})=0$. The pairing amplitudes are chosen to be $\Delta_s=1.88~\rm meV$, $\Delta_p=0.76~\rm meV$, and $\Delta_d=0.54~\rm meV$, to match the gap profile measured in the differential conductance with STM~\cite{27mingevidence2023}.

Back to real space, we consider a finite-size hexagonal geometry with each edge of the hexagonal containing $L=120$ unit cells. At the center of the hexagon, we consider a delta-function-like, non-magnetic potential impurity $h_1=U_0\delta(\boldsymbol{r})$ for the chiral $p$-wave and $d$-wave SCs. In the $s$-wave case, a magnetic impurity is required to break Cooper pairs and induce bound states, modeled by $h_1=(U_0+U_m\sigma_z)\delta(\boldsymbol{r})$. In all simulations, periodic boundary conditions are applied along all edges to eliminate edge states, ensuring any in-gap states solely arise from the impurity potential. Under the BdG framework, impurity-induced bound states appear in particle-hole conjugate pairs, denoted by $\Psi_{\alpha\pm}(\boldsymbol{r})$ at $\pm E_\alpha$, respectively. The positive-energy eigenstate can be expressed as $\Psi_{\alpha+}(\boldsymbol{r})=[u_\alpha(\boldsymbol{r}),v_\alpha(\boldsymbol{r})]^T$, where $u(\boldsymbol{r})$ and $v(\boldsymbol{r})$ are the electron and hole components of the Bogoliubov quasiparticles. Particle-hole symmetry imposes the relation $\Psi_{\alpha-}(\boldsymbol{r})=[v_\alpha^*(\boldsymbol{r}),u_\alpha^*(\boldsymbol{r})]^T$. The LDOS is computed as 
\begin{equation}
\rho(\omega, \boldsymbol{r}) = \sum_{\alpha} \left[ \delta(\omega - E_\alpha) |u_\alpha(\boldsymbol{r})|^2 + \delta(\omega + E_\alpha) |v_\alpha(\boldsymbol{r})|^2 \right],
\label{eq3}
\end{equation}
where the index $\alpha$ runs over all in-gap bound states.

\subsection{Origin of nodal impurity states}
We establish an analytical theory for nonmagnetic impurity states in a two-dimensional chiral $d$-wave superconductor. Consider a BdG Hamiltonian with a spin-singlet $d+id$ pairing,
\begin{equation}
H(\boldsymbol{k}) = 
\begin{bmatrix}
\epsilon_{\boldsymbol{k}} - \mu & \Delta_0\, k_+^2 \\
\Delta_0\, k_-^2 & \mu - \epsilon_{\boldsymbol{k}}
\end{bmatrix},
\end{equation}
under the Nambu basis $\Psi(\boldsymbol{k}) =( c_{\boldsymbol{k}, \uparrow}, c_{\boldsymbol{k}, \downarrow}, c^\dagger_{-\boldsymbol{k}, \downarrow}, -c^\dagger_{-\boldsymbol{k}, \uparrow})^T$. We have employed a different notation of $\Psi(\boldsymbol{k})$ compared to that of Eq.~(\ref{eq2}), with which $H(\boldsymbol{k})$ is decomposed into two decoupled yet identical matrix blocks, one for each spin sector. From now on, we will drop the spin indices for convenience. Without loss of generality, we consider an isotropic electron dispersion $\epsilon_{\boldsymbol{k}}=k^2/(2m)$, and further define $\xi_{\boldsymbol{k}}=\epsilon_{\boldsymbol{k}}-\mu$ for shorthand notation. The conclusion derived here remains robust even if we relax the isotropic condition and include the Rashba effect~\cite{44caiDeciphering2025}. We will now switch to the polar coordinates $(k_x,k_y )=k(\cos\theta,\sin\theta)$, where $k_\pm=ke^{\pm i\theta}$. The corresponding momentum-space Green’s function is thus given by
    \begin{eqnarray}
\mathcal{G}(\omega, \boldsymbol{k}) =  \frac{1}{\Omega(\boldsymbol{k},\omega)}\begin{bmatrix}
\omega + i\delta + \xi_{\boldsymbol{k}} & \Delta_0 k^2 e^{2i\theta} \\
\Delta_0 k^2 e^{-2i\theta} & \omega + i\delta - \xi_{\boldsymbol{k}}
\end{bmatrix},
\end{eqnarray}
where $\Omega(\boldsymbol{k},\omega) = (\omega + i\delta)^2 - \xi_{\boldsymbol{k}}^2 - \Delta_0^2 k^4$. We consider a point-like, non-magnetic impurity at $\boldsymbol{r}=0$, as described by a potential term $U_0 \tau_z \delta(\boldsymbol{r})$, where $\tau_{x,y,z}$ are Pauli matrices in the particle-hole space. The bound-state energy E and the bound-state wavefunction $\Psi(0)$ at the impurity site can be calculated by solving the following matrix equation~\cite{48PientkaTopological2013,49kaladzhyanasymptotic2016}:
\begin{equation}
\left[ \tau_0 - U_0 \tau_z \, \mathcal{G}(E, \boldsymbol{r} = 0) \right] \Psi(0) = 0.
\label{eq6}
\end{equation}
Here, the real-space Green’s function is given by $\mathcal{G}(E,\boldsymbol{r}=0)=\int \frac{d^2 \boldsymbol{k}}{(2\pi)^2} \, \mathcal{G}(E, \boldsymbol{k})$. While getting an exact analytical expression can be generally tricky, we note that $\mathcal{G}(E,\boldsymbol{r}=0)$ must be diagonal with a vanishing anomalous propagator part: 
\begin{widetext}
    \begin{equation}
\left[\mathcal{G}(E,\boldsymbol{r}=0)\right]_{12} = 
\int_0^{2\pi} \frac{dk}{(2\pi)^2} \, 
\frac{\Delta_0 k^3}{(E + i\delta)^2 - \xi_k^2 - \Delta_0^2 k^4} 
\int_0^{2\pi} e^{2i\theta} \, d\theta = 0.
\end{equation}
\end{widetext}
Physically, this implies a complete, local decoupling between electron and hole degrees of freedom at the impurity site, arising from the chirality of Cooper pairs. It is then easy to show that the eigenstate solutions for Eq.~(\ref{eq6}) must follow
\begin{equation}
\Psi_1(0) = 
\begin{pmatrix}
0 \\
1
\end{pmatrix}, \quad
\Psi_2(0) = 
\begin{pmatrix}
1 \\
0
\end{pmatrix},
\end{equation}
an inherent consequence of chiral Cooper pairs.

Summarizing, there must exist a pair of impurity-bound states at opposite energies: one with a vanishing electron component and the other with a vanishing hole component at the impurity site. Combined with the LDOS expression $\rho(\omega,\boldsymbol{r})$ in Eq.~(\ref{eq3}), we conclude that $\Psi_1(0)$ yields exactly zero LDOS at the impurity site at energy $E_0$, i.e., a nodal point, while the LDOS at $-E_0$, arising from $\Psi_2 (0)$, remains finite. This asymmetry explains the striking node–antinode structure seen in both simulation and experiment, which we identify as a robust and intrinsic consequence of chiral Cooper pairing.

\begin{figure}[t]
    \includegraphics[width=1.0\columnwidth]{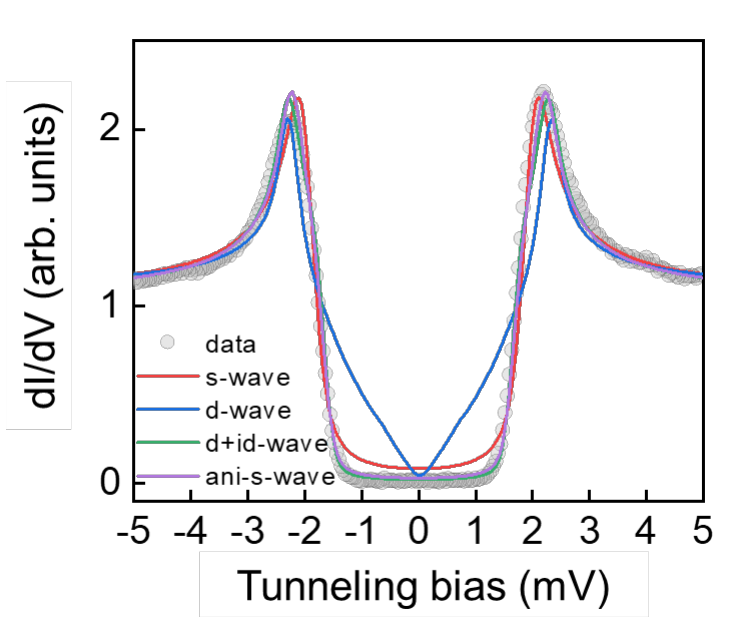}
    \caption{$dI/dV$ tunneling spectrum of the Sn/Si(111) system, recorded far from surface defects, along with theoretical fits for different superconducting order parameters (see text).} 
    \label{Extendfig1}
\end{figure}

\begin{figure}[t] 
    \centering
    \includegraphics[width=0.46\textwidth]{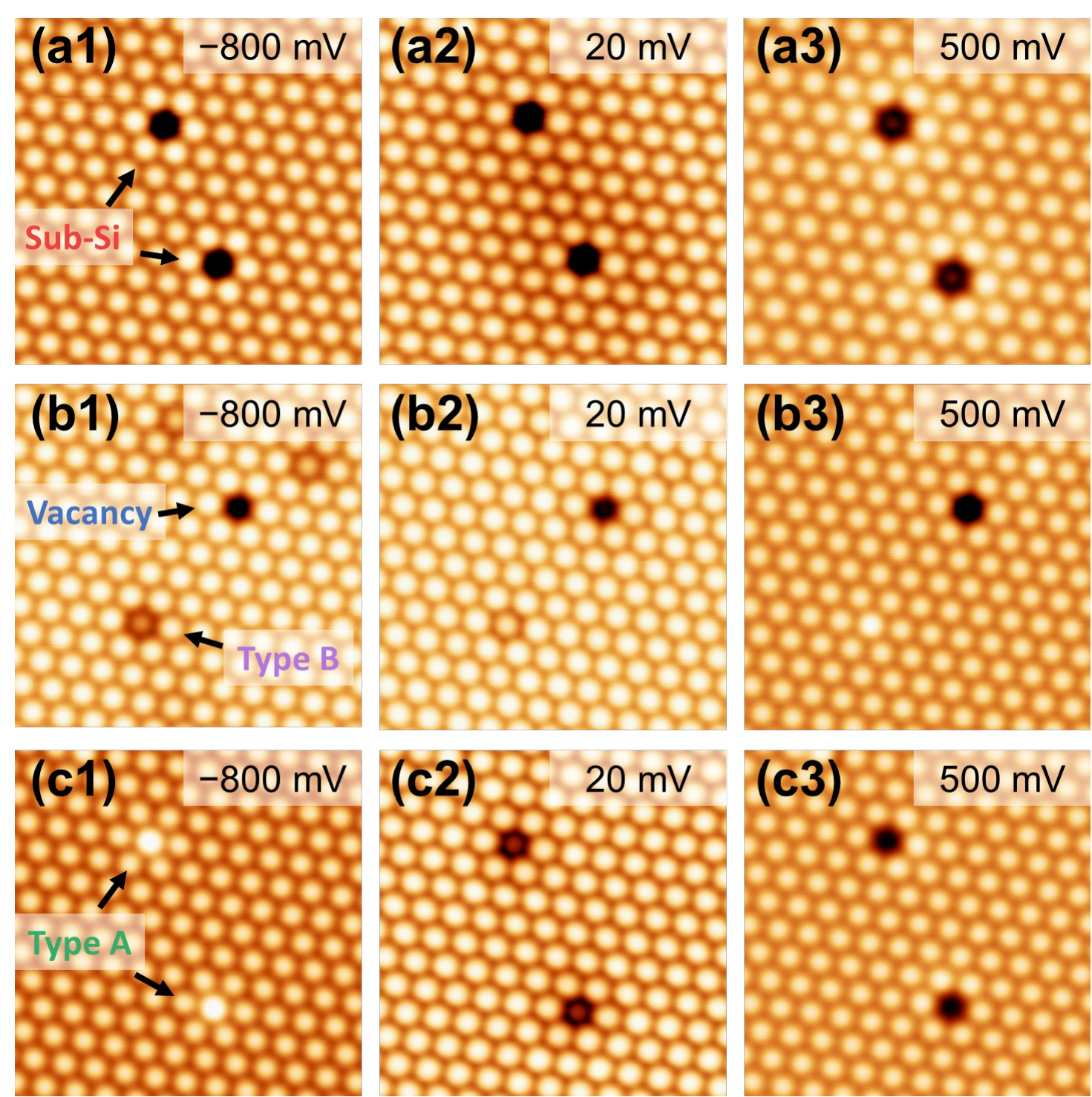}
    \caption{Bias dependent STM images of the various point defects in the Sn adatom layer. (a) Pair of substitutional Si defects, (b) a vacancy plus Type B defect, (c) a pair of Type A defects (See text). }
    \label{Extendfig2}
\end{figure}

\begin{figure}[t] 
    \centering
    \includegraphics[width=0.48\textwidth]{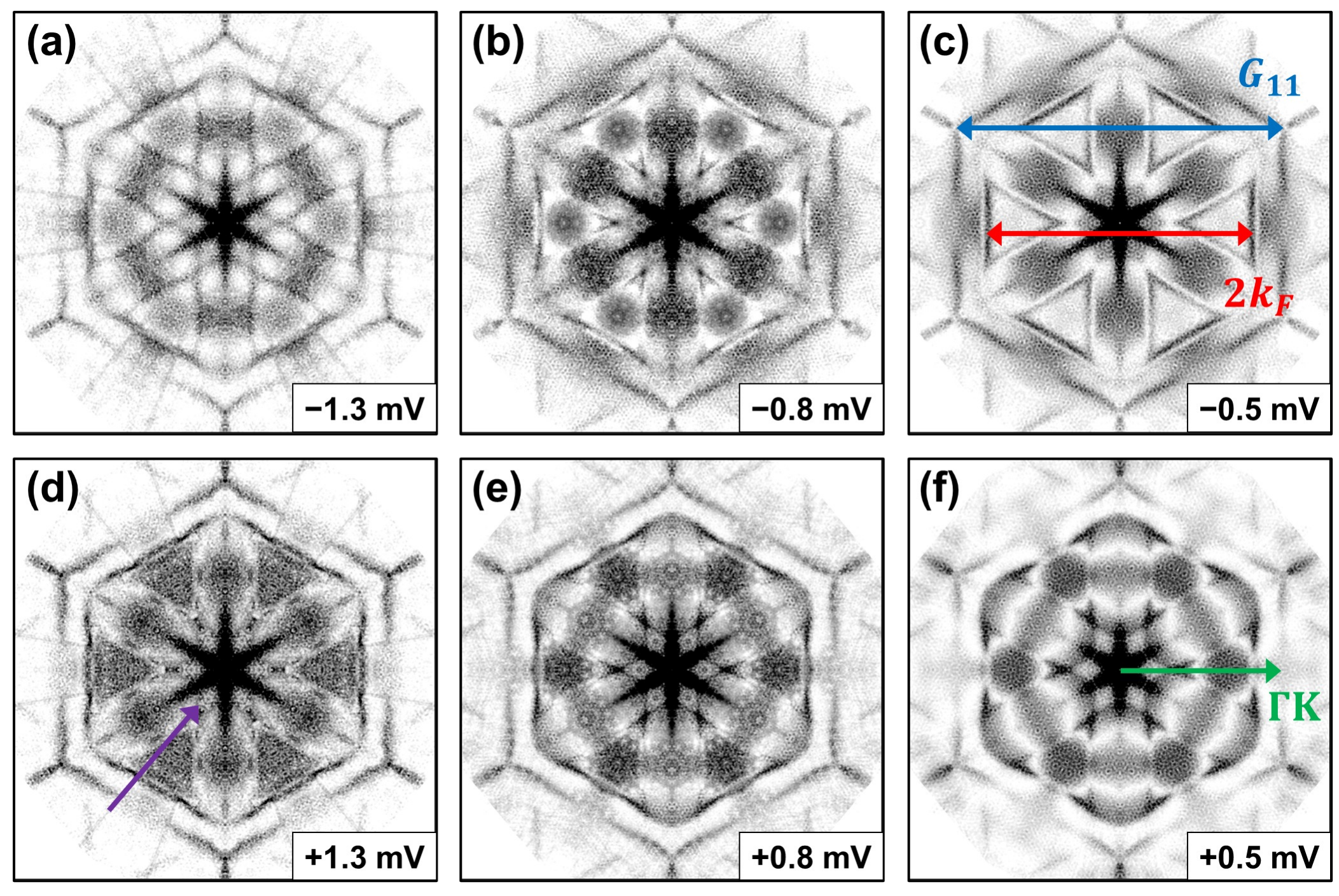}
    \caption{Fourier transformed QPI images for different tunnel voltages. The Fermi wavevector and $\boldsymbol{G}_{11}$ reciprocal lattice vector are indicated in panel (c). For a detailed explanation of these momentum space images, see Refs.~\cite{27mingevidence2023,31MingRealization2017}}.
    \label{Extendfig3}
\end{figure}

\section{Fitting the tunneling spectrum}

Figure~\ref{Extendfig1} shows a $dI/dV$ tunneling spectrum of the Sn/Si(111) system, recorded far from surface defects, along with theoretical fits for different superconducting order parameters. A detailed description of the fitting procedure can be found in Ref.~\cite{26WuXuefengSuperconductivity2020}. The conventional $d$-wave order parameter features a nodal gap and produces a V-shaped spectrum (blue line), which is inconsistent with the experimental data. Both the conventional $s$-wave (red) and chiral $d$-wave (green) order parameters yield a hard, U-shaped gap consistent with the measurements. The chiral $d$-wave order parameter clearly produces a better fit, yielding an optimized pairing amplitude of $\Delta_0=0.98~\rm meV$ and phenomenological broadening parameter $\Gamma=0.04~\rm meV$. A slightly better fit can be achieved using an anisotropic $s$-wave order parameter of the form $\Delta(\theta) = \Delta_0 + \Delta_1 \cos(2\theta)$, where $\Delta_0=1.94~\rm meV$ represents the isotropic gap component, and $\Delta_1=0.28~\rm meV$ captures the angular anisotropy of the pairing. Here, $\Delta(\theta)$ denotes the momentum-dependent gap function at the Fermi wavevector $k_\text{F}(\theta)$; $\theta$ is the azimuthal angle in $\boldsymbol{k}$-space. This fit introduces an additional fitting parameter ($\Delta_1$) and requires a slightly larger broadening $\Gamma=0.06~\rm meV$. It is worth noting that an anisotropic $s$-wave order parameter does not generate in-gap bound states in the presence of non-magnetic defects. A chiral $p$-wave fit is not shown, as it poorly reproduces the data~\cite{27mingevidence2023}. This failure occurs because any $p$-wave gap function must vanish at the $M$ point, which coincides with the location of the van Hove singularity at 7~meV below $E_\text{F}$.~\cite{27mingevidence2023}

\begin{figure*}[hbtp] 
    \centering
    \includegraphics[width=0.8\textwidth]{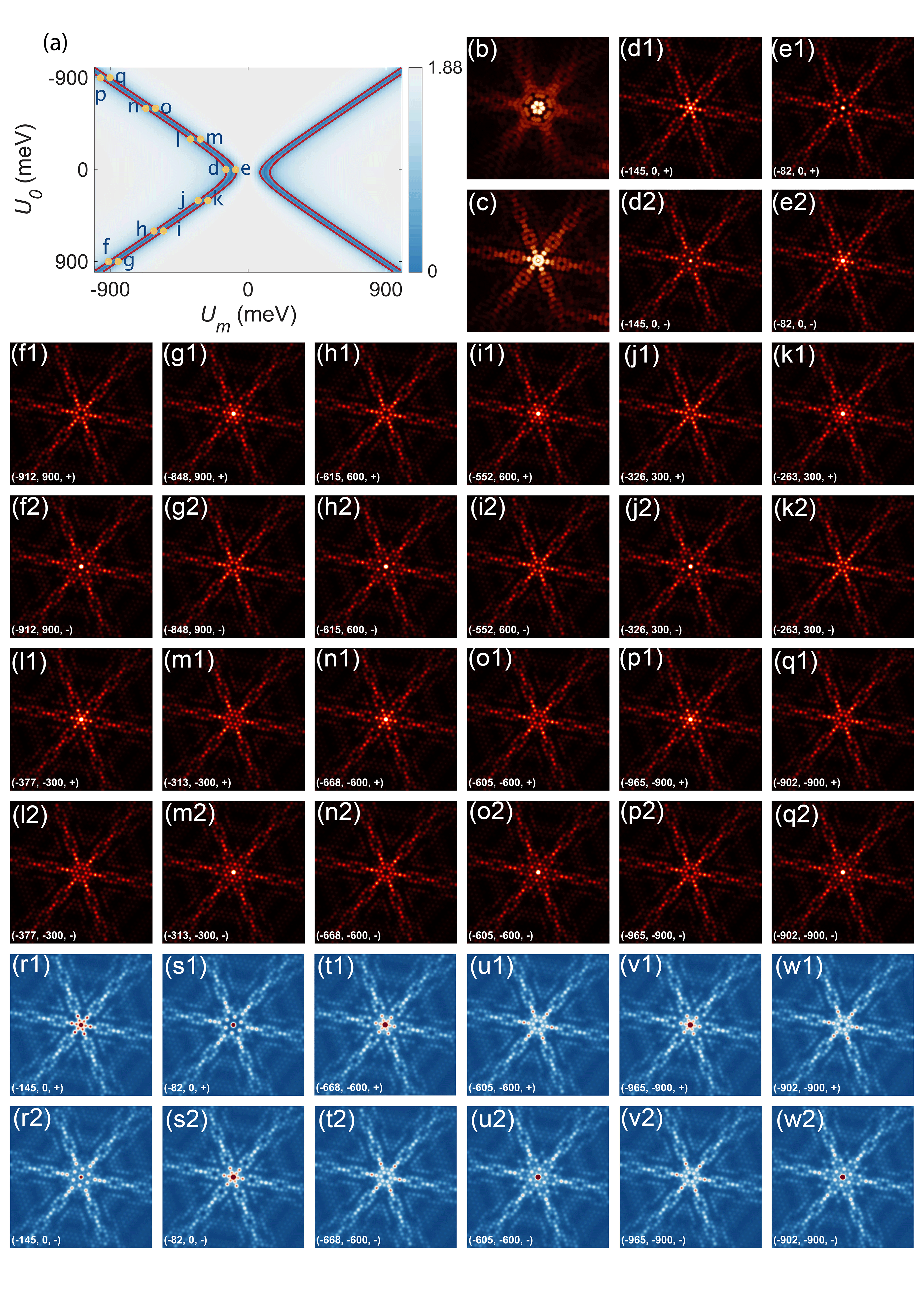}
    \caption{QPI image simulations for an $s$-wave order parameter. (a) Bound-state energy $E_0$ as a function of magnetic $U_m$ and non-magnetic $U_0$ impurity potential, as defined in the color bar. We take the pairing amplitude $\Delta_0=1.88$~meV. Red contours mark constant-energy lines at $E_0=0.54~\rm meV$ corresponding to the bound-state energy associated with a substitutional Si defect shown in Fig.~\ref{fig4}. (b) and (c) Experimental LDOS maps of a Si substitutional defect from Fig.~\ref{fig4} (d)-(q) Simulated LDOS maps following the red contours in (a). (r)-(w) Simulated LDOS maps for anisotropic $s$-wave pairing, using $\Delta_0=1.94~\rm meV$ and $\Delta_1=0.28~\rm meV$. For details, see text.}
    \label{Extendfig4}
\end{figure*}

\section{STM characterization of point defects}
As shown in Figure~\ref{Extendfig2}, STM images ($7.5\times7.5 ~\rm{nm}^2$) acquired at various tunneling biases (with constant tunneling current $I_t = 0.1~\rm nA$) show three surface regions containing point defects. Four distinct types of point defects are labeled in panels (a1), (b1), and (c1). Among them, the substitutional Si defect and the Sn adatom vacancy are identified based on their bias-dependent STM signatures, following Refs.~~\cite{27mingevidence2023} and ~\cite{50jemanderstm2001}. The substitutional Si defect (labeled “Sub-Si” in panels (a1)–(a3)) appears as a vacancy at a sample bias of $V_s = -800~\rm mV$, with its six nearest-neighbor Sn adatoms slightly brighter than the surrounding ones. At $V_s = 500~\rm mV$, the Si adatom becomes visible, while nearby Sn adatoms appear somewhat darker at lower biases such as $V_s = -20~\rm mV$. The Sn adatom vacancy (top defect in panels (b1)–(b3)) consistently appears as a missing Sn adatom at all biases, with minimal contrast changes in the surrounding adatoms relative to the pristine Sn background. Two additional point defects, labeled Type A and Type B, both exhibiting $C_6$ symmetry, are frequently observed and show distinct bias-dependent contrasts. The Type A defect (panels (c1)–(c3)) appears as a bright adatom at $V_s = -800~\rm mV$ but becomes dim in filled-state images at $V_s = 20~\rm mV$ and $500~\rm mV$. In contrast, the Type B defect (lower defect in panels (b1)–(b3)) appears dim at $V_s = -800~\rm mV$ and $20~\rm mV$, but bright at $V_s = 500~\rm mV$. Type A and Type B defects are absent in Sn/Si(111) samples grown on n-type Si wafers doped with As (data not shown), suggesting they are associated with (subsurface) B atoms, although their exact atomic structures remain undetermined.

\section{Momentum space QPI}

Figure~\ref{Extendfig3} shows momentum-space QPI images of Sn/Si(111), recorded at bias voltages within the superconducting gap. The images are similar to those of Ref.~~\cite{27mingevidence2023} but are better resolved. Panel (c) highlights the scattering vectors $2\boldsymbol{k}_\text{F}$  and $\boldsymbol{G}_{11}$, where $\boldsymbol{G}_{11}$ connects the nearest Bragg points along the $\overline{\Gamma K}$ direction in momentum space. The Fermi wavevector magnitude was determined using the Bragg peaks for scale calibration. The $\overline{\Gamma K}$ direction is indicated in panel (f). The arrow in panel (d) points to a flower-like feature centered at the origin of momentum space, which arises from scattering processes that break time-reversal symmetry~\cite{27mingevidence2023}. It is replicated at the other Bragg positions.

\section{Additional Impurity State Simulations}

Figure~\ref{Extendfig4} presents additional impurity state simulations of Sn/Si(111) with $s$-wave superconductivity. (a) plots the bound-state energy $E_0$ as a function of magnetic $U_m$ and non-magnetic $U_0$ impurity potential, as defined in the color bar. Red contours mark constant-energy lines at $E_0=0.54~\rm meV$ corresponding to the bound-state energy associated with a substitutional Si defect shown in Fig.~\ref{fig4}. Along these contours, we compute impurity-induced LDOS for 82 distinct $(U_m,U_0)$ combinations, each yielding two LDOS maps at $\pm E_0$. Due to space constraints, only 14 representative cases (orange dots), labeled as (d) through (q), are shown. For each point X, the simulated LDOS at $+E_0$ and $-E_0$ are displayed in panels (X1) and (X2), respectively, with annotations indicating the potential parameters and bound-state energies $(U_m,U_0,\pm)$  in meV. (b) and (c), Experimental LDOS maps of a Si substitutional defect at $\pm0.54$ mV, shown for comparison. Numerically, we find that flipping the sign of $U_m\rightarrow-U_m$ leaves both the bound-state energy and LDOS unchanged, so we restrict our simulations to $U_m<0$. Panels (r1)-(w2) show LDOS calculations with an anisotropic $s$-wave pairing, where the extended $s$-wave pairing component is $\Delta_1=0.28~\rm meV$, following the gap fitting. These panels use a distinct color scale to differentiate them from the conventional $s$-wave results. The following panels share the same impurity parameters: (d) and (r), (e) and (s), (n) and (t), (o) and (u), (p) and (v), and (q) and (w). The resulting LDOS patterns are nearly indistinguishable, differing only by minor quantitative details. Crucially, none of the $s$-wave simulations—whether conventional or anisotropic—reproduce the experimentally observed flower-star pattern (see panels (b) and (c)).

\begin{figure}[htbp]
    \includegraphics[width=1.0\columnwidth]{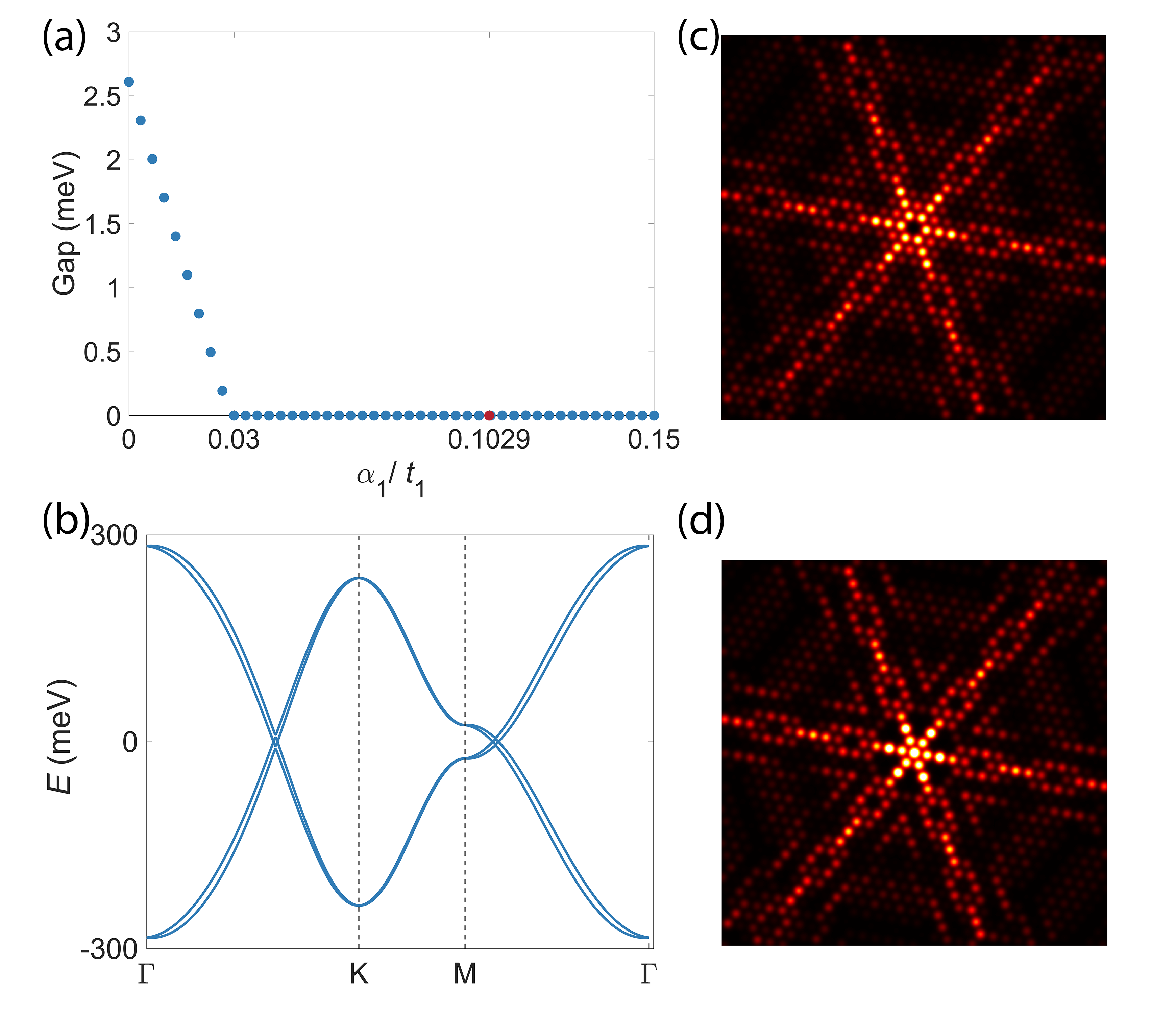}
    \caption{Impurity state simulations of Sn/Si(111) for a chiral $p$-wave order parameter, including Rashba spin-orbital coupling. (a) BdG gap as a function of Rashba spin-orbital coupling $\alpha_1$. (b) momentum-resolved BdG spectrum with $\alpha_1=0.1029t_1$. (c) and (d) Simulated LDOS maps at $+0.54~\rm meV$  and $-0.54~\rm meV$ with $\alpha_1=0$, respectively. See text.} 
    \label{Extendfig5}
\end{figure}

Figure~\ref{Extendfig5} shows impurity state simulations of Sn/Si(111) with chiral $p$-wave superconductivity. Panel (a) plots the evolution of BdG energy gap as a function of Rashba spin-orbital coupling $\alpha_1$, showing a transition from nodeless to nodal behavior for $\alpha_1\geq0.03t_1$. The gap closing arises from the competition between Rashba effect and spin-triplet $p$-wave pairing. Panel (b) shows the momentum-resolved BdG spectrum with the ab-initio Rashba parameter $\alpha_1=0.1029t_1$, revealing gapless BdG bands caused by the Rashba spin-splitting. Simulated LDOS maps at $+0.54~\rm meV$  and $-0.54~\rm meV$ with $\alpha_1=0$ are shown in panels (c) and (d), respectively. Similar to the chiral $d$-wave case, the node-antinode structure at the impurity site reflects the chiral nature of pairing order. However, the overall patterns fail to reproduce the experimental features.

\section{Broken rotational symmetry}

\begin{figure}[htbp] 
    \centering
    \includegraphics[width=0.47\textwidth]{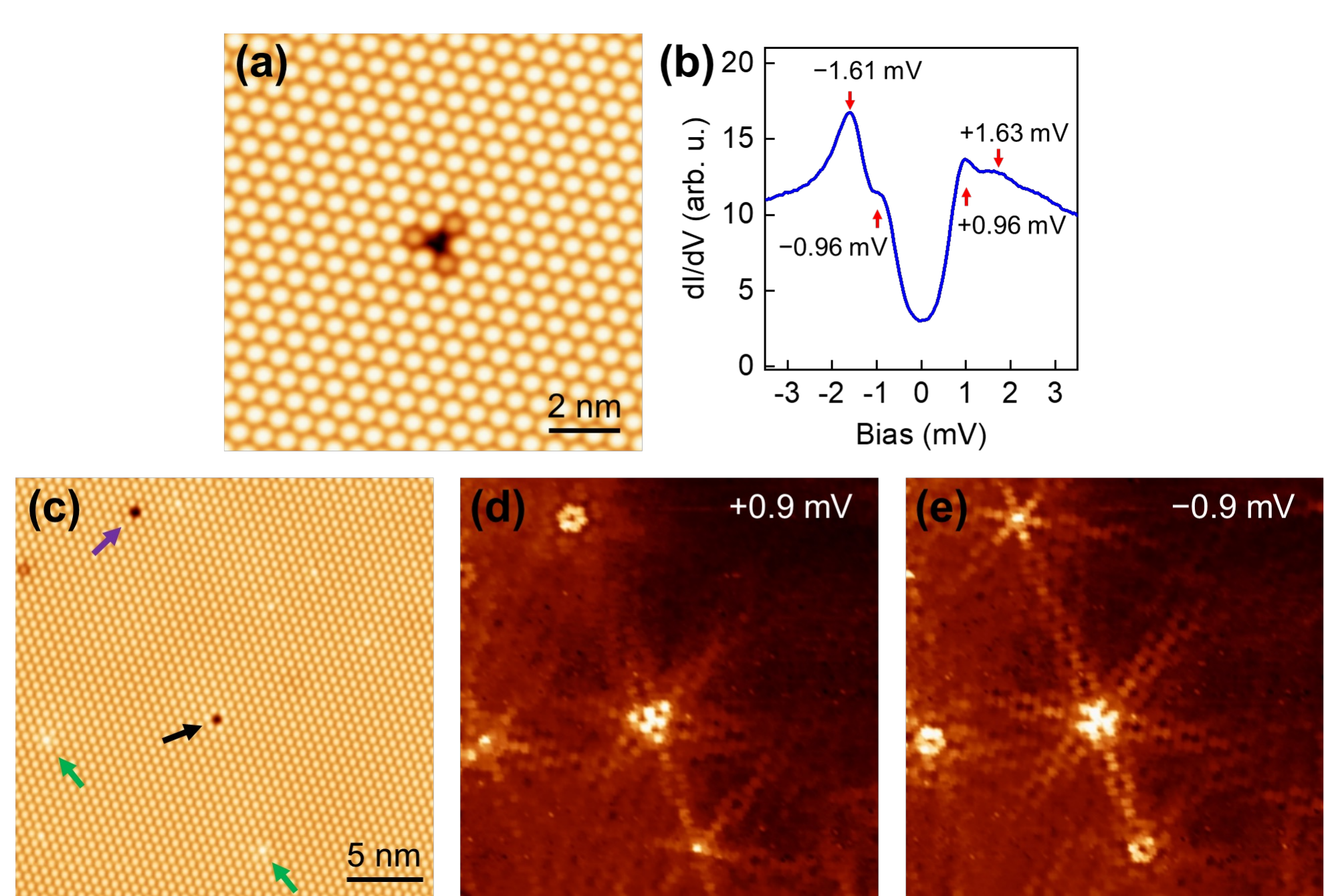}
    \caption{(a) STM image ($V_s=20~\rm mV$, $I_t=0.1~\rm nA$) of a man-made vacancy, created by removing a Sn adatom via tip-induced manipulation using $V_s=-2.0~\rm V$ in a defect-free region. 
    (b) $dI/dV$ spectrum acquired on top of the man-made vacancy, revealing several in-gap states. (c)  STM image ($V_s=-2.0~\rm V$, $I_t=1~\rm nA$) of the same region, highlighting the man-made vacancy (black arrow), one Type B defect (purple arrow), and two Type A defects (green arrows). (d), (e) $dI/dV$ maps of the same region as in panel (c). 
    }
    \label{Extendfig6}
\end{figure}

Figure~\ref{Extendfig6} shows QPI images near a man-made vacancy defect. Panel (a) provides an STM image ($V_s=20~\rm mV$, $I_t=0.1~\rm nA$) of a man-made vacancy, created by removing a Sn adatom via tip-induced manipulation using $V_s=-2.0~\rm V$ in a defect-free region. The resulting vacancy, located at the original Sn adatom site, appears similar to a natural vacancy in STM topography. However, at low bias, three of the six nearest-neighbor Sn adatoms appear dimmer, indicating a structural difference from natural vacancies. We speculate that this reflects a frozen metastable configuration of a natural vacancy, stabilized at low temperature. Panel (b) shows a $dI/dV$ spectrum acquired on top of the man-made vacancy, revealing several in-gap states. Panel (c) shows an STM image ($V_s=-2.0~\rm V$, $I_t=1~\rm nA$) of the same region, highlighting the man-made vacancy (black arrow), one Type B defect (purple arrow), and two Type A defects (green arrows). Panels (d) and (e) show $dI/dV$ maps of the same region as in panel (c). Notably, the man-made vacancy lacks the star-like or hexagon-like spatial features at opposite bias polarities that are characteristic of Type A and Type B defects. They also do not show the distinct node-antinode feature at the defect location. Evidently, the chiral signature is hidden due to the increased complexity of the (threefold) scattering potential.

\bibliography{ref}

\begin{thebibliography}{50}%
\makeatletter
\providecommand \@ifxundefined [1]{%
 \@ifx{#1\undefined}
}%
\providecommand \@ifnum [1]{%
 \ifnum #1\expandafter \@firstoftwo
 \else \expandafter \@secondoftwo
 \fi
}%
\providecommand \@ifx [1]{%
 \ifx #1\expandafter \@firstoftwo
 \else \expandafter \@secondoftwo
 \fi
}%
\providecommand \natexlab [1]{#1}%
\providecommand \enquote  [1]{``#1''}%
\providecommand \bibnamefont  [1]{#1}%
\providecommand \bibfnamefont [1]{#1}%
\providecommand \citenamefont [1]{#1}%
\providecommand \href@noop [0]{\@secondoftwo}%
\providecommand \href [0]{\begingroup \@sanitize@url \@href}%
\providecommand \@href[1]{\@@startlink{#1}\@@href}%
\providecommand \@@href[1]{\endgroup#1\@@endlink}%
\providecommand \@sanitize@url [0]{\catcode `\\12\catcode `\$12\catcode
  `\&12\catcode `\#12\catcode `\^12\catcode `\_12\catcode `\%12\relax}%
\providecommand \@@startlink[1]{}%
\providecommand \@@endlink[0]{}%
\providecommand \url  [0]{\begingroup\@sanitize@url \@url }%
\providecommand \@url [1]{\endgroup\@href {#1}{\urlprefix }}%
\providecommand \urlprefix  [0]{URL }%
\providecommand \Eprint [0]{\href }%
\providecommand \doibase [0]{https://doi.org/}%
\providecommand \selectlanguage [0]{\@gobble}%
\providecommand \bibinfo  [0]{\@secondoftwo}%
\providecommand \bibfield  [0]{\@secondoftwo}%
\providecommand \translation [1]{[#1]}%
\providecommand \BibitemOpen [0]{}%
\providecommand \bibitemStop [0]{}%
\providecommand \bibitemNoStop [0]{.\EOS\space}%
\providecommand \EOS [0]{\spacefactor3000\relax}%
\providecommand \BibitemShut  [1]{\csname bibitem#1\endcsname}%
\let\auto@bib@innerbib\@empty
\bibitem [{\citenamefont {Read}\ and\ \citenamefont
  {Green}(2000)}]{1ReadPaired2000}%
  \BibitemOpen
  \bibfield  {author} {\bibinfo {author} {\bibfnamefont {N.}~\bibnamefont
  {Read}}\ and\ \bibinfo {author} {\bibfnamefont {D.}~\bibnamefont {Green}},\
  }\bibfield  {title} {\bibinfo {title} {Paired states of fermions in two
  dimensions with breaking of parity and time-reversal symmetries and the
  fractional quantum hall effect},\ }\href
  {https://link.aps.org/doi/10.1103/PhysRevB.61.10267} {\bibfield  {journal}
  {\bibinfo  {journal} {Phys. Rev. B}\ }\textbf {\bibinfo {volume} {61}},\
  \bibinfo {pages} {10267} (\bibinfo {year} {2000})}\BibitemShut {NoStop}%
\bibitem [{\citenamefont {Nayak}\ \emph {et~al.}(2008)\citenamefont {Nayak},
  \citenamefont {Simon}, \citenamefont {Stern}, \citenamefont {Freedman},\ and\
  \citenamefont {Das~Sarma}}]{2SankarNon-Abelian2008}%
  \BibitemOpen
  \bibfield  {author} {\bibinfo {author} {\bibfnamefont {C.}~\bibnamefont
  {Nayak}}, \bibinfo {author} {\bibfnamefont {S.~H.}\ \bibnamefont {Simon}},
  \bibinfo {author} {\bibfnamefont {A.}~\bibnamefont {Stern}}, \bibinfo
  {author} {\bibfnamefont {M.}~\bibnamefont {Freedman}},\ and\ \bibinfo
  {author} {\bibfnamefont {S.}~\bibnamefont {Das~Sarma}},\ }\bibfield  {title}
  {\bibinfo {title} {Non-abelian anyons and topological quantum computation},\
  }\href {https://link.aps.org/doi/10.1103/RevModPhys.80.1083} {\bibfield
  {journal} {\bibinfo  {journal} {Rev. Mod. Phys.}\ }\textbf {\bibinfo {volume}
  {80}},\ \bibinfo {pages} {1083} (\bibinfo {year} {2008})}\BibitemShut
  {NoStop}%
\bibitem [{\citenamefont {Sato}\ and\ \citenamefont
  {Ando}(2017)}]{3satotopological2017}%
  \BibitemOpen
  \bibfield  {author} {\bibinfo {author} {\bibfnamefont {M.}~\bibnamefont
  {Sato}}\ and\ \bibinfo {author} {\bibfnamefont {Y.}~\bibnamefont {Ando}},\
  }\bibfield  {title} {\bibinfo {title} {Topological superconductors: a
  review},\ }\href {https://dx.doi.org/10.1088/1361-6633/aa6ac7} {\bibfield
  {journal} {\bibinfo  {journal} {Reports on Progress in Physics}\ }\textbf
  {\bibinfo {volume} {80}},\ \bibinfo {pages} {076501} (\bibinfo {year}
  {2017})}\BibitemShut {NoStop}%
\bibitem [{\citenamefont {Luke}\ \emph {et~al.}(1998)\citenamefont {Luke},
  \citenamefont {Fudamoto}, \citenamefont {Kojima}, \citenamefont {Larkin},
  \citenamefont {Merrin}, \citenamefont {Nachumi}, \citenamefont {Uemura},
  \citenamefont {Maeno}, \citenamefont {Mao}, \citenamefont {Mori},
  \citenamefont {Nakamura},\ and\ \citenamefont
  {Sigrist}}]{4luketime-reversal1998}%
  \BibitemOpen
  \bibfield  {author} {\bibinfo {author} {\bibfnamefont {G.~M.}\ \bibnamefont
  {Luke}}, \bibinfo {author} {\bibfnamefont {Y.}~\bibnamefont {Fudamoto}},
  \bibinfo {author} {\bibfnamefont {K.~M.}\ \bibnamefont {Kojima}}, \bibinfo
  {author} {\bibfnamefont {M.~I.}\ \bibnamefont {Larkin}}, \bibinfo {author}
  {\bibfnamefont {J.}~\bibnamefont {Merrin}}, \bibinfo {author} {\bibfnamefont
  {B.}~\bibnamefont {Nachumi}}, \bibinfo {author} {\bibfnamefont {Y.~J.}\
  \bibnamefont {Uemura}}, \bibinfo {author} {\bibfnamefont {Y.}~\bibnamefont
  {Maeno}}, \bibinfo {author} {\bibfnamefont {Z.~Q.}\ \bibnamefont {Mao}},
  \bibinfo {author} {\bibfnamefont {Y.}~\bibnamefont {Mori}}, \bibinfo {author}
  {\bibfnamefont {H.}~\bibnamefont {Nakamura}},\ and\ \bibinfo {author}
  {\bibfnamefont {M.}~\bibnamefont {Sigrist}},\ }\bibfield  {title} {\bibinfo
  {title} {Time-reversal symmetry-breaking superconductivity in sr2ruo4},\
  }\href {https://doi.org/10.1038/29038} {\bibfield  {journal} {\bibinfo
  {journal} {Nature}\ }\textbf {\bibinfo {volume} {394}},\ \bibinfo {pages}
  {558} (\bibinfo {year} {1998})}\BibitemShut {NoStop}%
\bibitem [{\citenamefont {Mackenzie}\ and\ \citenamefont
  {Maeno}(2003)}]{5MackenzieThesuperconductivity2003}%
  \BibitemOpen
  \bibfield  {author} {\bibinfo {author} {\bibfnamefont {A.~P.}\ \bibnamefont
  {Mackenzie}}\ and\ \bibinfo {author} {\bibfnamefont {Y.}~\bibnamefont
  {Maeno}},\ }\bibfield  {title} {\bibinfo {title} {The superconductivity of
  ${\mathrm{sr}}_{2}{\mathrm{ruo}}_{4}$ and the physics of spin-triplet
  pairing},\ }\href {https://link.aps.org/doi/10.1103/RevModPhys.75.657}
  {\bibfield  {journal} {\bibinfo  {journal} {Rev. Mod. Phys.}\ }\textbf
  {\bibinfo {volume} {75}},\ \bibinfo {pages} {657} (\bibinfo {year}
  {2003})}\BibitemShut {NoStop}%
\bibitem [{\citenamefont {Xia}\ \emph {et~al.}(2006)\citenamefont {Xia},
  \citenamefont {Maeno}, \citenamefont {Beyersdorf}, \citenamefont {Fejer},\
  and\ \citenamefont {Kapitulnik}}]{6XiaHigh2006}%
  \BibitemOpen
  \bibfield  {author} {\bibinfo {author} {\bibfnamefont {J.}~\bibnamefont
  {Xia}}, \bibinfo {author} {\bibfnamefont {Y.}~\bibnamefont {Maeno}}, \bibinfo
  {author} {\bibfnamefont {P.~T.}\ \bibnamefont {Beyersdorf}}, \bibinfo
  {author} {\bibfnamefont {M.~M.}\ \bibnamefont {Fejer}},\ and\ \bibinfo
  {author} {\bibfnamefont {A.}~\bibnamefont {Kapitulnik}},\ }\bibfield  {title}
  {\bibinfo {title} {High resolution polar kerr effect measurements of
  ${\mathrm{sr}}_{2}{\mathrm{ruo}}_{4}$: Evidence for broken time-reversal
  symmetry in the superconducting state},\ }\href
  {https://link.aps.org/doi/10.1103/PhysRevLett.97.167002} {\bibfield
  {journal} {\bibinfo  {journal} {Phys. Rev. Lett.}\ }\textbf {\bibinfo
  {volume} {97}},\ \bibinfo {pages} {167002} (\bibinfo {year}
  {2006})}\BibitemShut {NoStop}%
\bibitem [{\citenamefont {Pustogow}\ \emph {et~al.}(2019)\citenamefont
  {Pustogow}, \citenamefont {Luo}, \citenamefont {Chronister}, \citenamefont
  {Su}, \citenamefont {Sokolov}, \citenamefont {Jerzembeck}, \citenamefont
  {Mackenzie}, \citenamefont {Hicks}, \citenamefont {Kikugawa}, \citenamefont
  {Raghu}, \citenamefont {Bauer},\ and\ \citenamefont
  {Brown}}]{7pustogowconstraints2019}%
  \BibitemOpen
  \bibfield  {author} {\bibinfo {author} {\bibfnamefont {A.}~\bibnamefont
  {Pustogow}}, \bibinfo {author} {\bibfnamefont {Y.}~\bibnamefont {Luo}},
  \bibinfo {author} {\bibfnamefont {A.}~\bibnamefont {Chronister}}, \bibinfo
  {author} {\bibfnamefont {Y.-S.}\ \bibnamefont {Su}}, \bibinfo {author}
  {\bibfnamefont {D.~A.}\ \bibnamefont {Sokolov}}, \bibinfo {author}
  {\bibfnamefont {F.}~\bibnamefont {Jerzembeck}}, \bibinfo {author}
  {\bibfnamefont {A.~P.}\ \bibnamefont {Mackenzie}}, \bibinfo {author}
  {\bibfnamefont {C.~W.}\ \bibnamefont {Hicks}}, \bibinfo {author}
  {\bibfnamefont {N.}~\bibnamefont {Kikugawa}}, \bibinfo {author}
  {\bibfnamefont {S.}~\bibnamefont {Raghu}}, \bibinfo {author} {\bibfnamefont
  {E.~D.}\ \bibnamefont {Bauer}},\ and\ \bibinfo {author} {\bibfnamefont
  {S.~E.}\ \bibnamefont {Brown}},\ }\bibfield  {title} {\bibinfo {title}
  {Constraints on the superconducting order parameter in sr2ruo4 from oxygen-17
  nuclear magnetic resonance},\ }\href
  {https://doi.org/10.1038/s41586-019-1596-2} {\bibfield  {journal} {\bibinfo
  {journal} {Nature}\ }\textbf {\bibinfo {volume} {574}},\ \bibinfo {pages}
  {72} (\bibinfo {year} {2019})}\BibitemShut {NoStop}%
\bibitem [{\citenamefont {Chronister}\ \emph {et~al.}(2021)\citenamefont
  {Chronister}, \citenamefont {Pustogow}, \citenamefont {Kikugawa},
  \citenamefont {Sokolov}, \citenamefont {Jerzembeck}, \citenamefont {Hicks},
  \citenamefont {Mackenzie}, \citenamefont {Bauer},\ and\ \citenamefont
  {Brown}}]{8AaronEvidence2021}%
  \BibitemOpen
  \bibfield  {author} {\bibinfo {author} {\bibfnamefont {A.}~\bibnamefont
  {Chronister}}, \bibinfo {author} {\bibfnamefont {A.}~\bibnamefont
  {Pustogow}}, \bibinfo {author} {\bibfnamefont {N.}~\bibnamefont {Kikugawa}},
  \bibinfo {author} {\bibfnamefont {D.~A.}\ \bibnamefont {Sokolov}}, \bibinfo
  {author} {\bibfnamefont {F.}~\bibnamefont {Jerzembeck}}, \bibinfo {author}
  {\bibfnamefont {C.~W.}\ \bibnamefont {Hicks}}, \bibinfo {author}
  {\bibfnamefont {A.~P.}\ \bibnamefont {Mackenzie}}, \bibinfo {author}
  {\bibfnamefont {E.~D.}\ \bibnamefont {Bauer}},\ and\ \bibinfo {author}
  {\bibfnamefont {S.~E.}\ \bibnamefont {Brown}},\ }\bibfield  {title} {\bibinfo
  {title} {Evidence for even parity unconventional superconductivity in
  sr$_2$ruo$_4$},\ }\href
  {https://www.pnas.org/doi/abs/10.1073/pnas.2025313118} {\bibfield  {journal}
  {\bibinfo  {journal} {Proceedings of the National Academy of Sciences}\
  }\textbf {\bibinfo {volume} {118}},\ \bibinfo {pages} {e2025313118} (\bibinfo
  {year} {2021})}\BibitemShut {NoStop}%
\bibitem [{\citenamefont {Aoki}\ \emph {et~al.}(2019)\citenamefont {Aoki},
  \citenamefont {Nakamura}, \citenamefont {Honda}, \citenamefont {Li},
  \citenamefont {Homma}, \citenamefont {Shimizu}, \citenamefont {Sato},
  \citenamefont {Knebel}, \citenamefont {Brison}, \citenamefont {Pourret},
  \citenamefont {Braithwaite}, \citenamefont {Lapertot}, \citenamefont {Niu},
  \citenamefont {Vališka}, \citenamefont {Harima},\ and\ \citenamefont
  {Flouquet}}]{9aokiunconventional2019}%
  \BibitemOpen
  \bibfield  {author} {\bibinfo {author} {\bibfnamefont {D.}~\bibnamefont
  {Aoki}}, \bibinfo {author} {\bibfnamefont {A.}~\bibnamefont {Nakamura}},
  \bibinfo {author} {\bibfnamefont {F.}~\bibnamefont {Honda}}, \bibinfo
  {author} {\bibfnamefont {D.}~\bibnamefont {Li}}, \bibinfo {author}
  {\bibfnamefont {Y.}~\bibnamefont {Homma}}, \bibinfo {author} {\bibfnamefont
  {Y.}~\bibnamefont {Shimizu}}, \bibinfo {author} {\bibfnamefont {Y.~J.}\
  \bibnamefont {Sato}}, \bibinfo {author} {\bibfnamefont {G.}~\bibnamefont
  {Knebel}}, \bibinfo {author} {\bibfnamefont {J.-P.}\ \bibnamefont {Brison}},
  \bibinfo {author} {\bibfnamefont {A.}~\bibnamefont {Pourret}}, \bibinfo
  {author} {\bibfnamefont {D.}~\bibnamefont {Braithwaite}}, \bibinfo {author}
  {\bibfnamefont {G.}~\bibnamefont {Lapertot}}, \bibinfo {author}
  {\bibfnamefont {Q.}~\bibnamefont {Niu}}, \bibinfo {author} {\bibfnamefont
  {M.}~\bibnamefont {Vališka}}, \bibinfo {author} {\bibfnamefont
  {H.}~\bibnamefont {Harima}},\ and\ \bibinfo {author} {\bibfnamefont
  {J.}~\bibnamefont {Flouquet}},\ }\bibfield  {title} {\bibinfo {title}
  {Unconventional superconductivity in heavy fermion ute$_2$},\ }\href
  {https://doi.org/10.7566/JPSJ.88.043702} {\bibfield  {journal} {\bibinfo
  {journal} {Journal of the Physical Society of Japan}\ }\textbf {\bibinfo
  {volume} {88}},\ \bibinfo {pages} {043702} (\bibinfo {year}
  {2019})}\BibitemShut {NoStop}%
\bibitem [{\citenamefont {Ran}\ \emph {et~al.}(2019)\citenamefont {Ran},
  \citenamefont {Eckberg}, \citenamefont {Ding}, \citenamefont {Furukawa},
  \citenamefont {Metz}, \citenamefont {Saha}, \citenamefont {Liu},
  \citenamefont {Zic}, \citenamefont {Kim}, \citenamefont {Paglione},\ and\
  \citenamefont {Butch}}]{10ShengNearly2019}%
  \BibitemOpen
  \bibfield  {author} {\bibinfo {author} {\bibfnamefont {S.}~\bibnamefont
  {Ran}}, \bibinfo {author} {\bibfnamefont {C.}~\bibnamefont {Eckberg}},
  \bibinfo {author} {\bibfnamefont {Q.-P.}\ \bibnamefont {Ding}}, \bibinfo
  {author} {\bibfnamefont {Y.}~\bibnamefont {Furukawa}}, \bibinfo {author}
  {\bibfnamefont {T.}~\bibnamefont {Metz}}, \bibinfo {author} {\bibfnamefont
  {S.~R.}\ \bibnamefont {Saha}}, \bibinfo {author} {\bibfnamefont {I.-L.}\
  \bibnamefont {Liu}}, \bibinfo {author} {\bibfnamefont {M.}~\bibnamefont
  {Zic}}, \bibinfo {author} {\bibfnamefont {H.}~\bibnamefont {Kim}}, \bibinfo
  {author} {\bibfnamefont {J.}~\bibnamefont {Paglione}},\ and\ \bibinfo
  {author} {\bibfnamefont {N.~P.}\ \bibnamefont {Butch}},\ }\bibfield  {title}
  {\bibinfo {title} {Nearly ferromagnetic spin-triplet superconductivity},\
  }\href {https://www.science.org/doi/abs/10.1126/science.aav8645} {\bibfield
  {journal} {\bibinfo  {journal} {Science}\ }\textbf {\bibinfo {volume}
  {365}},\ \bibinfo {pages} {684} (\bibinfo {year} {2019})}\BibitemShut
  {NoStop}%
\bibitem [{\citenamefont {Gu}\ \emph {et~al.}(2025)\citenamefont {Gu},
  \citenamefont {Wang}, \citenamefont {Carroll}, \citenamefont {Zhussupbekov},
  \citenamefont {Broyles}, \citenamefont {Ran}, \citenamefont {Butch},
  \citenamefont {Horn}, \citenamefont {Saha}, \citenamefont {Paglione},
  \citenamefont {Liu}, \citenamefont {Davis},\ and\ \citenamefont
  {Lee}}]{11QiangqiangPair2025}%
  \BibitemOpen
  \bibfield  {author} {\bibinfo {author} {\bibfnamefont {Q.}~\bibnamefont
  {Gu}}, \bibinfo {author} {\bibfnamefont {S.}~\bibnamefont {Wang}}, \bibinfo
  {author} {\bibfnamefont {J.~P.}\ \bibnamefont {Carroll}}, \bibinfo {author}
  {\bibfnamefont {K.}~\bibnamefont {Zhussupbekov}}, \bibinfo {author}
  {\bibfnamefont {C.}~\bibnamefont {Broyles}}, \bibinfo {author} {\bibfnamefont
  {S.}~\bibnamefont {Ran}}, \bibinfo {author} {\bibfnamefont {N.~P.}\
  \bibnamefont {Butch}}, \bibinfo {author} {\bibfnamefont {J.~A.}\ \bibnamefont
  {Horn}}, \bibinfo {author} {\bibfnamefont {S.}~\bibnamefont {Saha}}, \bibinfo
  {author} {\bibfnamefont {J.}~\bibnamefont {Paglione}}, \bibinfo {author}
  {\bibfnamefont {X.}~\bibnamefont {Liu}}, \bibinfo {author} {\bibfnamefont
  {J.~C.~S.}\ \bibnamefont {Davis}},\ and\ \bibinfo {author} {\bibfnamefont
  {D.-H.}\ \bibnamefont {Lee}},\ }\bibfield  {title} {\bibinfo {title} {Pair
  wave function symmetry in ute$_2$ from zero-energy surface state
  visualization},\ }\href
  {https://www.science.org/doi/abs/10.1126/science.adk7219} {\bibfield
  {journal} {\bibinfo  {journal} {Science}\ }\textbf {\bibinfo {volume}
  {388}},\ \bibinfo {pages} {938} (\bibinfo {year} {2025})}\BibitemShut
  {NoStop}%
\bibitem [{\citenamefont {Schemm}\ \emph {et~al.}(2014)\citenamefont {Schemm},
  \citenamefont {Gannon}, \citenamefont {Wishne}, \citenamefont {Halperin},\
  and\ \citenamefont {Kapitulnik}}]{12ERSchemmObservation2014}%
  \BibitemOpen
  \bibfield  {author} {\bibinfo {author} {\bibfnamefont {E.~R.}\ \bibnamefont
  {Schemm}}, \bibinfo {author} {\bibfnamefont {W.~J.}\ \bibnamefont {Gannon}},
  \bibinfo {author} {\bibfnamefont {C.~M.}\ \bibnamefont {Wishne}}, \bibinfo
  {author} {\bibfnamefont {W.~P.}\ \bibnamefont {Halperin}},\ and\ \bibinfo
  {author} {\bibfnamefont {A.}~\bibnamefont {Kapitulnik}},\ }\bibfield  {title}
  {\bibinfo {title} {Observation of broken time-reversal symmetry in the
  heavy-fermion superconductor upt3},\ }\href
  {https://www.science.org/doi/abs/10.1126/science.1248552} {\bibfield
  {journal} {\bibinfo  {journal} {Science}\ }\textbf {\bibinfo {volume}
  {345}},\ \bibinfo {pages} {190} (\bibinfo {year} {2014})}\BibitemShut
  {NoStop}%
\bibitem [{\citenamefont {Avers}\ \emph {et~al.}(2020)\citenamefont {Avers},
  \citenamefont {Gannon}, \citenamefont {Kuhn}, \citenamefont {Halperin},
  \citenamefont {Sauls}, \citenamefont {{DeBeer}-Schmitt}, \citenamefont
  {Dewhurst}, \citenamefont {Gavilano}, \citenamefont {Nagy}, \citenamefont
  {Gasser},\ and\ \citenamefont {Eskildsen}}]{13aversbroken2020}%
  \BibitemOpen
  \bibfield  {author} {\bibinfo {author} {\bibfnamefont {K.~E.}\ \bibnamefont
  {Avers}}, \bibinfo {author} {\bibfnamefont {W.~J.}\ \bibnamefont {Gannon}},
  \bibinfo {author} {\bibfnamefont {S.~J.}\ \bibnamefont {Kuhn}}, \bibinfo
  {author} {\bibfnamefont {W.~P.}\ \bibnamefont {Halperin}}, \bibinfo {author}
  {\bibfnamefont {J.~A.}\ \bibnamefont {Sauls}}, \bibinfo {author}
  {\bibfnamefont {L.}~\bibnamefont {{DeBeer}-Schmitt}}, \bibinfo {author}
  {\bibfnamefont {C.~D.}\ \bibnamefont {Dewhurst}}, \bibinfo {author}
  {\bibfnamefont {J.}~\bibnamefont {Gavilano}}, \bibinfo {author}
  {\bibfnamefont {G.}~\bibnamefont {Nagy}}, \bibinfo {author} {\bibfnamefont
  {U.}~\bibnamefont {Gasser}},\ and\ \bibinfo {author} {\bibfnamefont {M.~R.}\
  \bibnamefont {Eskildsen}},\ }\bibfield  {title} {\bibinfo {title} {Broken
  time-reversal symmetry in the topological superconductor upt3},\ }\href
  {https://doi.org/10.1038/s41567-020-0822-z} {\bibfield  {journal} {\bibinfo
  {journal} {Nature Physics}\ }\textbf {\bibinfo {volume} {16}},\ \bibinfo
  {pages} {531} (\bibinfo {year} {2020})}\BibitemShut {NoStop}%
\bibitem [{\citenamefont {Ribak}\ \emph {et~al.}(2020)\citenamefont {Ribak},
  \citenamefont {Skiff}, \citenamefont {Mograbi}, \citenamefont {Rout},
  \citenamefont {Fischer}, \citenamefont {Ruhman}, \citenamefont {Chashka},
  \citenamefont {Dagan},\ and\ \citenamefont {Kanigel}}]{14ARibakChiral2020}%
  \BibitemOpen
  \bibfield  {author} {\bibinfo {author} {\bibfnamefont {A.}~\bibnamefont
  {Ribak}}, \bibinfo {author} {\bibfnamefont {R.~M.}\ \bibnamefont {Skiff}},
  \bibinfo {author} {\bibfnamefont {M.}~\bibnamefont {Mograbi}}, \bibinfo
  {author} {\bibfnamefont {P.~K.}\ \bibnamefont {Rout}}, \bibinfo {author}
  {\bibfnamefont {M.~H.}\ \bibnamefont {Fischer}}, \bibinfo {author}
  {\bibfnamefont {J.}~\bibnamefont {Ruhman}}, \bibinfo {author} {\bibfnamefont
  {K.}~\bibnamefont {Chashka}}, \bibinfo {author} {\bibfnamefont
  {Y.}~\bibnamefont {Dagan}},\ and\ \bibinfo {author} {\bibfnamefont
  {A.}~\bibnamefont {Kanigel}},\ }\bibfield  {title} {\bibinfo {title} {Chiral
  superconductivity in the alternate stacking compound 4hb-tas2},\ }\href
  {https://www.science.org/doi/abs/10.1126/sciadv.aax9480} {\bibfield
  {journal} {\bibinfo  {journal} {Science Advances}\ }\textbf {\bibinfo
  {volume} {6}},\ \bibinfo {pages} {eaax9480} (\bibinfo {year}
  {2020})}\BibitemShut {NoStop}%
\bibitem [{\citenamefont {Zhao}\ \emph {et~al.}(2023)\citenamefont {Zhao},
  \citenamefont {Cui}, \citenamefont {Volkov}, \citenamefont {Yoo},
  \citenamefont {Lee}, \citenamefont {Gardener}, \citenamefont {Akey},
  \citenamefont {Engelke}, \citenamefont {Ronen}, \citenamefont {Zhong},
  \citenamefont {Gu}, \citenamefont {Plugge}, \citenamefont {Tummuru},
  \citenamefont {Kim}, \citenamefont {Franz}, \citenamefont {Pixley},
  \citenamefont {Poccia},\ and\ \citenamefont
  {Kim}}]{15SYFrankZhaoTime-reversal2023}%
  \BibitemOpen
  \bibfield  {author} {\bibinfo {author} {\bibfnamefont {S.~Y.~F.}\
  \bibnamefont {Zhao}}, \bibinfo {author} {\bibfnamefont {X.}~\bibnamefont
  {Cui}}, \bibinfo {author} {\bibfnamefont {P.~A.}\ \bibnamefont {Volkov}},
  \bibinfo {author} {\bibfnamefont {H.}~\bibnamefont {Yoo}}, \bibinfo {author}
  {\bibfnamefont {S.}~\bibnamefont {Lee}}, \bibinfo {author} {\bibfnamefont
  {J.~A.}\ \bibnamefont {Gardener}}, \bibinfo {author} {\bibfnamefont {A.~J.}\
  \bibnamefont {Akey}}, \bibinfo {author} {\bibfnamefont {R.}~\bibnamefont
  {Engelke}}, \bibinfo {author} {\bibfnamefont {Y.}~\bibnamefont {Ronen}},
  \bibinfo {author} {\bibfnamefont {R.}~\bibnamefont {Zhong}}, \bibinfo
  {author} {\bibfnamefont {G.}~\bibnamefont {Gu}}, \bibinfo {author}
  {\bibfnamefont {S.}~\bibnamefont {Plugge}}, \bibinfo {author} {\bibfnamefont
  {T.}~\bibnamefont {Tummuru}}, \bibinfo {author} {\bibfnamefont
  {M.}~\bibnamefont {Kim}}, \bibinfo {author} {\bibfnamefont {M.}~\bibnamefont
  {Franz}}, \bibinfo {author} {\bibfnamefont {J.~H.}\ \bibnamefont {Pixley}},
  \bibinfo {author} {\bibfnamefont {N.}~\bibnamefont {Poccia}},\ and\ \bibinfo
  {author} {\bibfnamefont {P.}~\bibnamefont {Kim}},\ }\bibfield  {title}
  {\bibinfo {title} {Time-reversal symmetry breaking superconductivity between
  twisted cuprate superconductors},\ }\href
  {https://www.science.org/doi/abs/10.1126/science.abl8371} {\bibfield
  {journal} {\bibinfo  {journal} {Science}\ }\textbf {\bibinfo {volume}
  {382}},\ \bibinfo {pages} {1422} (\bibinfo {year} {2023})}\BibitemShut
  {NoStop}%
\bibitem [{\citenamefont {Wan}\ \emph {et~al.}(2024)\citenamefont {Wan},
  \citenamefont {Qiu}, \citenamefont {Ren}, \citenamefont {Qian}, \citenamefont
  {Li}, \citenamefont {Xu}, \citenamefont {Zhou}, \citenamefont {Zhou},
  \citenamefont {Zhou}, \citenamefont {Wang}, \citenamefont {Yang},
  \citenamefont {Sofer}, \citenamefont {Huang}, \citenamefont {Wang},\ and\
  \citenamefont {Duan}}]{16wanunconventional2024}%
  \BibitemOpen
  \bibfield  {author} {\bibinfo {author} {\bibfnamefont {Z.}~\bibnamefont
  {Wan}}, \bibinfo {author} {\bibfnamefont {G.}~\bibnamefont {Qiu}}, \bibinfo
  {author} {\bibfnamefont {H.}~\bibnamefont {Ren}}, \bibinfo {author}
  {\bibfnamefont {Q.}~\bibnamefont {Qian}}, \bibinfo {author} {\bibfnamefont
  {Y.}~\bibnamefont {Li}}, \bibinfo {author} {\bibfnamefont {D.}~\bibnamefont
  {Xu}}, \bibinfo {author} {\bibfnamefont {J.}~\bibnamefont {Zhou}}, \bibinfo
  {author} {\bibfnamefont {J.}~\bibnamefont {Zhou}}, \bibinfo {author}
  {\bibfnamefont {B.}~\bibnamefont {Zhou}}, \bibinfo {author} {\bibfnamefont
  {L.}~\bibnamefont {Wang}}, \bibinfo {author} {\bibfnamefont {T.-H.}\
  \bibnamefont {Yang}}, \bibinfo {author} {\bibfnamefont {Z.}~\bibnamefont
  {Sofer}}, \bibinfo {author} {\bibfnamefont {Y.}~\bibnamefont {Huang}},
  \bibinfo {author} {\bibfnamefont {K.~L.}\ \bibnamefont {Wang}},\ and\
  \bibinfo {author} {\bibfnamefont {X.}~\bibnamefont {Duan}},\ }\bibfield
  {title} {\bibinfo {title} {Unconventional superconductivity in chiral
  molecule–tas2 hybrid superlattices},\ }\href
  {https://doi.org/10.1038/s41586-024-07625-4} {\bibfield  {journal} {\bibinfo
  {journal} {Nature}\ }\textbf {\bibinfo {volume} {632}},\ \bibinfo {pages}
  {69} (\bibinfo {year} {2024})}\BibitemShut {NoStop}%
\bibitem [{\citenamefont {Han}\ \emph {et~al.}(2025)\citenamefont {Han},
  \citenamefont {Lu}, \citenamefont {Hadjri}, \citenamefont {Shi},
  \citenamefont {Wu}, \citenamefont {Xu}, \citenamefont {Yao}, \citenamefont
  {Cotten}, \citenamefont {Sharifi~Sedeh}, \citenamefont {Weldeyesus},
  \citenamefont {Yang}, \citenamefont {Seo}, \citenamefont {Ye}, \citenamefont
  {Zhou}, \citenamefont {Liu}, \citenamefont {Shi}, \citenamefont {Hua},
  \citenamefont {Watanabe}, \citenamefont {Taniguchi}, \citenamefont {Xiong},
  \citenamefont {Zumbühl}, \citenamefont {Fu},\ and\ \citenamefont
  {Ju}}]{17hansignatures2025}%
  \BibitemOpen
  \bibfield  {author} {\bibinfo {author} {\bibfnamefont {T.}~\bibnamefont
  {Han}}, \bibinfo {author} {\bibfnamefont {Z.}~\bibnamefont {Lu}}, \bibinfo
  {author} {\bibfnamefont {Z.}~\bibnamefont {Hadjri}}, \bibinfo {author}
  {\bibfnamefont {L.}~\bibnamefont {Shi}}, \bibinfo {author} {\bibfnamefont
  {Z.}~\bibnamefont {Wu}}, \bibinfo {author} {\bibfnamefont {W.}~\bibnamefont
  {Xu}}, \bibinfo {author} {\bibfnamefont {Y.}~\bibnamefont {Yao}}, \bibinfo
  {author} {\bibfnamefont {A.~A.}\ \bibnamefont {Cotten}}, \bibinfo {author}
  {\bibfnamefont {O.}~\bibnamefont {Sharifi~Sedeh}}, \bibinfo {author}
  {\bibfnamefont {H.}~\bibnamefont {Weldeyesus}}, \bibinfo {author}
  {\bibfnamefont {J.}~\bibnamefont {Yang}}, \bibinfo {author} {\bibfnamefont
  {J.}~\bibnamefont {Seo}}, \bibinfo {author} {\bibfnamefont {S.}~\bibnamefont
  {Ye}}, \bibinfo {author} {\bibfnamefont {M.}~\bibnamefont {Zhou}}, \bibinfo
  {author} {\bibfnamefont {H.}~\bibnamefont {Liu}}, \bibinfo {author}
  {\bibfnamefont {G.}~\bibnamefont {Shi}}, \bibinfo {author} {\bibfnamefont
  {Z.}~\bibnamefont {Hua}}, \bibinfo {author} {\bibfnamefont {K.}~\bibnamefont
  {Watanabe}}, \bibinfo {author} {\bibfnamefont {T.}~\bibnamefont {Taniguchi}},
  \bibinfo {author} {\bibfnamefont {P.}~\bibnamefont {Xiong}}, \bibinfo
  {author} {\bibfnamefont {D.~M.}\ \bibnamefont {Zumbühl}}, \bibinfo {author}
  {\bibfnamefont {L.}~\bibnamefont {Fu}},\ and\ \bibinfo {author}
  {\bibfnamefont {L.}~\bibnamefont {Ju}},\ }\bibfield  {title} {\bibinfo
  {title} {Signatures of chiral superconductivity in rhombohedral graphene},\
  }\href {https://doi.org/10.1038/s41586-025-09169-7} {\bibfield  {journal}
  {\bibinfo  {journal} {Nature}\ }\textbf {\bibinfo {volume} {643}},\ \bibinfo
  {pages} {654} (\bibinfo {year} {2025})}\BibitemShut {NoStop}%
\bibitem [{\citenamefont {Soumyanarayanan}\ \emph {et~al.}(2013)\citenamefont
  {Soumyanarayanan}, \citenamefont {Yee}, \citenamefont {He}, \citenamefont
  {van Wezel}, \citenamefont {Rahn}, \citenamefont {Rossnagel}, \citenamefont
  {Hudson}, \citenamefont {Norman},\ and\ \citenamefont
  {Hoffman}}]{18AnjanQuantum2013}%
  \BibitemOpen
  \bibfield  {author} {\bibinfo {author} {\bibfnamefont {A.}~\bibnamefont
  {Soumyanarayanan}}, \bibinfo {author} {\bibfnamefont {M.~M.}\ \bibnamefont
  {Yee}}, \bibinfo {author} {\bibfnamefont {Y.}~\bibnamefont {He}}, \bibinfo
  {author} {\bibfnamefont {J.}~\bibnamefont {van Wezel}}, \bibinfo {author}
  {\bibfnamefont {D.~J.}\ \bibnamefont {Rahn}}, \bibinfo {author}
  {\bibfnamefont {K.}~\bibnamefont {Rossnagel}}, \bibinfo {author}
  {\bibfnamefont {E.~W.}\ \bibnamefont {Hudson}}, \bibinfo {author}
  {\bibfnamefont {M.~R.}\ \bibnamefont {Norman}},\ and\ \bibinfo {author}
  {\bibfnamefont {J.~E.}\ \bibnamefont {Hoffman}},\ }\bibfield  {title}
  {\bibinfo {title} {Quantum phase transition from triangular to stripe charge
  order in nbse2},\ }\href
  {https://www.pnas.org/doi/abs/10.1073/pnas.1211387110} {\bibfield  {journal}
  {\bibinfo  {journal} {Proceedings of the National Academy of Sciences}\
  }\textbf {\bibinfo {volume} {110}},\ \bibinfo {pages} {1623} (\bibinfo {year}
  {2013})}\BibitemShut {NoStop}%
\bibitem [{\citenamefont {Wang}\ \emph {et~al.}(2020)\citenamefont {Wang},
  \citenamefont {Rodriguez}, \citenamefont {Jiao}, \citenamefont {Howard},
  \citenamefont {Graham}, \citenamefont {Gu}, \citenamefont {Hughes},
  \citenamefont {Morr},\ and\ \citenamefont {Madhavan}}]{19ZhenyuEvidence2020}%
  \BibitemOpen
  \bibfield  {author} {\bibinfo {author} {\bibfnamefont {Z.}~\bibnamefont
  {Wang}}, \bibinfo {author} {\bibfnamefont {J.~O.}\ \bibnamefont {Rodriguez}},
  \bibinfo {author} {\bibfnamefont {L.}~\bibnamefont {Jiao}}, \bibinfo {author}
  {\bibfnamefont {S.}~\bibnamefont {Howard}}, \bibinfo {author} {\bibfnamefont
  {M.}~\bibnamefont {Graham}}, \bibinfo {author} {\bibfnamefont {G.~D.}\
  \bibnamefont {Gu}}, \bibinfo {author} {\bibfnamefont {T.~L.}\ \bibnamefont
  {Hughes}}, \bibinfo {author} {\bibfnamefont {D.~K.}\ \bibnamefont {Morr}},\
  and\ \bibinfo {author} {\bibfnamefont {V.}~\bibnamefont {Madhavan}},\
  }\bibfield  {title} {\bibinfo {title} {Evidence for dispersing 1d majorana
  channels in an iron-based superconductor},\ }\href
  {https://www.science.org/doi/abs/10.1126/science.aaw8419} {\bibfield
  {journal} {\bibinfo  {journal} {Science}\ }\textbf {\bibinfo {volume}
  {367}},\ \bibinfo {pages} {104} (\bibinfo {year} {2020})}\BibitemShut
  {NoStop}%
\bibitem [{\citenamefont {Hanaguri}\ \emph {et~al.}(2010)\citenamefont
  {Hanaguri}, \citenamefont {Niitaka}, \citenamefont {Kuroki},\ and\
  \citenamefont {Takagi}}]{20THanaguriUnconventional2010}%
  \BibitemOpen
  \bibfield  {author} {\bibinfo {author} {\bibfnamefont {T.}~\bibnamefont
  {Hanaguri}}, \bibinfo {author} {\bibfnamefont {S.}~\bibnamefont {Niitaka}},
  \bibinfo {author} {\bibfnamefont {K.}~\bibnamefont {Kuroki}},\ and\ \bibinfo
  {author} {\bibfnamefont {H.}~\bibnamefont {Takagi}},\ }\bibfield  {title}
  {\bibinfo {title} {Unconventional s-wave superconductivity in fe(se,te)},\
  }\href {https://www.science.org/doi/abs/10.1126/science.1187399} {\bibfield
  {journal} {\bibinfo  {journal} {Science}\ }\textbf {\bibinfo {volume}
  {328}},\ \bibinfo {pages} {474} (\bibinfo {year} {2010})}\BibitemShut
  {NoStop}%
\bibitem [{\citenamefont {McElroy}\ \emph {et~al.}(2005)\citenamefont
  {McElroy}, \citenamefont {Lee}, \citenamefont {Slezak}, \citenamefont {Lee},
  \citenamefont {Eisaki}, \citenamefont {Uchida},\ and\ \citenamefont
  {Davis}}]{21KMcElroyAtomic-Scale2005}%
  \BibitemOpen
  \bibfield  {author} {\bibinfo {author} {\bibfnamefont {K.}~\bibnamefont
  {McElroy}}, \bibinfo {author} {\bibfnamefont {J.}~\bibnamefont {Lee}},
  \bibinfo {author} {\bibfnamefont {J.~A.}\ \bibnamefont {Slezak}}, \bibinfo
  {author} {\bibfnamefont {D.-H.}\ \bibnamefont {Lee}}, \bibinfo {author}
  {\bibfnamefont {H.}~\bibnamefont {Eisaki}}, \bibinfo {author} {\bibfnamefont
  {S.}~\bibnamefont {Uchida}},\ and\ \bibinfo {author} {\bibfnamefont {J.~C.}\
  \bibnamefont {Davis}},\ }\bibfield  {title} {\bibinfo {title} {Atomic-scale
  sources and mechanism of nanoscale electronic disorder in
  bi2sr2cacu2o8+$\delta$},\ }\href
  {https://www.science.org/doi/abs/10.1126/science.1113095} {\bibfield
  {journal} {\bibinfo  {journal} {Science}\ }\textbf {\bibinfo {volume}
  {309}},\ \bibinfo {pages} {1048} (\bibinfo {year} {2005})}\BibitemShut
  {NoStop}%
\bibitem [{\citenamefont {Allan}\ \emph {et~al.}(2013)\citenamefont {Allan},
  \citenamefont {Massee}, \citenamefont {Morr}, \citenamefont {Van~Dyke},
  \citenamefont {Rost}, \citenamefont {Mackenzie}, \citenamefont {Petrovic},\
  and\ \citenamefont {Davis}}]{22allanimaging2013}%
  \BibitemOpen
  \bibfield  {author} {\bibinfo {author} {\bibfnamefont {M.~P.}\ \bibnamefont
  {Allan}}, \bibinfo {author} {\bibfnamefont {F.}~\bibnamefont {Massee}},
  \bibinfo {author} {\bibfnamefont {D.~K.}\ \bibnamefont {Morr}}, \bibinfo
  {author} {\bibfnamefont {J.}~\bibnamefont {Van~Dyke}}, \bibinfo {author}
  {\bibfnamefont {A.~W.}\ \bibnamefont {Rost}}, \bibinfo {author}
  {\bibfnamefont {A.~P.}\ \bibnamefont {Mackenzie}}, \bibinfo {author}
  {\bibfnamefont {C.}~\bibnamefont {Petrovic}},\ and\ \bibinfo {author}
  {\bibfnamefont {J.~C.}\ \bibnamefont {Davis}},\ }\bibfield  {title} {\bibinfo
  {title} {Imaging cooper pairing of heavy fermions in cecoin5},\ }\href
  {https://doi.org/10.1038/nphys2671} {\bibfield  {journal} {\bibinfo
  {journal} {Nature Physics}\ }\textbf {\bibinfo {volume} {9}},\ \bibinfo
  {pages} {468} (\bibinfo {year} {2013})}\BibitemShut {NoStop}%
\bibitem [{\citenamefont {Gao}\ \emph {et~al.}(2018)\citenamefont {Gao},
  \citenamefont {Wang}, \citenamefont {Zhou}, \citenamefont {Huang},\ and\
  \citenamefont {Wang}}]{23GaoPossible2018}%
  \BibitemOpen
  \bibfield  {author} {\bibinfo {author} {\bibfnamefont {Y.}~\bibnamefont
  {Gao}}, \bibinfo {author} {\bibfnamefont {Y.}~\bibnamefont {Wang}}, \bibinfo
  {author} {\bibfnamefont {T.}~\bibnamefont {Zhou}}, \bibinfo {author}
  {\bibfnamefont {H.}~\bibnamefont {Huang}},\ and\ \bibinfo {author}
  {\bibfnamefont {Q.-H.}\ \bibnamefont {Wang}},\ }\bibfield  {title} {\bibinfo
  {title} {Possible pairing symmetry in the fese-based superconductors
  determined by quasiparticle interference},\ }\href
  {https://link.aps.org/doi/10.1103/PhysRevLett.121.267005} {\bibfield
  {journal} {\bibinfo  {journal} {Phys. Rev. Lett.}\ }\textbf {\bibinfo
  {volume} {121}},\ \bibinfo {pages} {267005} (\bibinfo {year}
  {2018})}\BibitemShut {NoStop}%
\bibitem [{\citenamefont {Böker}\ \emph {et~al.}(2019)\citenamefont {Böker},
  \citenamefont {Volkov}, \citenamefont {Hirschfeld},\ and\ \citenamefont
  {Eremin}}]{24bokerquasiparticle2019}%
  \BibitemOpen
  \bibfield  {author} {\bibinfo {author} {\bibfnamefont {J.}~\bibnamefont
  {Böker}}, \bibinfo {author} {\bibfnamefont {P.~A.}\ \bibnamefont {Volkov}},
  \bibinfo {author} {\bibfnamefont {P.~J.}\ \bibnamefont {Hirschfeld}},\ and\
  \bibinfo {author} {\bibfnamefont {I.}~\bibnamefont {Eremin}},\ }\bibfield
  {title} {\bibinfo {title} {Quasiparticle interference and symmetry of
  superconducting order parameter in strongly electron-doped iron-based
  superconductors},\ }\href {https://dx.doi.org/10.1088/1367-2630/ab2a82}
  {\bibfield  {journal} {\bibinfo  {journal} {New Journal of Physics}\ }\textbf
  {\bibinfo {volume} {21}},\ \bibinfo {pages} {083021} (\bibinfo {year}
  {2019})}\BibitemShut {NoStop}%
\bibitem [{\citenamefont {Kreisel}\ \emph {et~al.}(2015)\citenamefont
  {Kreisel}, \citenamefont {Choubey}, \citenamefont {Berlijn}, \citenamefont
  {Ku}, \citenamefont {Andersen},\ and\ \citenamefont
  {Hirschfeld}}]{25KreiselInterpretation2015}%
  \BibitemOpen
  \bibfield  {author} {\bibinfo {author} {\bibfnamefont {A.}~\bibnamefont
  {Kreisel}}, \bibinfo {author} {\bibfnamefont {P.}~\bibnamefont {Choubey}},
  \bibinfo {author} {\bibfnamefont {T.}~\bibnamefont {Berlijn}}, \bibinfo
  {author} {\bibfnamefont {W.}~\bibnamefont {Ku}}, \bibinfo {author}
  {\bibfnamefont {B.~M.}\ \bibnamefont {Andersen}},\ and\ \bibinfo {author}
  {\bibfnamefont {P.~J.}\ \bibnamefont {Hirschfeld}},\ }\bibfield  {title}
  {\bibinfo {title} {Interpretation of scanning tunneling quasiparticle
  interference and impurity states in cuprates},\ }\href
  {https://link.aps.org/doi/10.1103/PhysRevLett.114.217002} {\bibfield
  {journal} {\bibinfo  {journal} {Phys. Rev. Lett.}\ }\textbf {\bibinfo
  {volume} {114}},\ \bibinfo {pages} {217002} (\bibinfo {year}
  {2015})}\BibitemShut {NoStop}%
\bibitem [{\citenamefont {Ming}\ \emph {et~al.}(2017)\citenamefont {Ming},
  \citenamefont {Johnston}, \citenamefont {Mulugeta}, \citenamefont {Smith},
  \citenamefont {Vilmercati}, \citenamefont {Lee}, \citenamefont {Maier},
  \citenamefont {Snijders},\ and\ \citenamefont
  {Weitering}}]{31MingRealization2017}%
  \BibitemOpen
  \bibfield  {author} {\bibinfo {author} {\bibfnamefont {F.}~\bibnamefont
  {Ming}}, \bibinfo {author} {\bibfnamefont {S.}~\bibnamefont {Johnston}},
  \bibinfo {author} {\bibfnamefont {D.}~\bibnamefont {Mulugeta}}, \bibinfo
  {author} {\bibfnamefont {T.~S.}\ \bibnamefont {Smith}}, \bibinfo {author}
  {\bibfnamefont {P.}~\bibnamefont {Vilmercati}}, \bibinfo {author}
  {\bibfnamefont {G.}~\bibnamefont {Lee}}, \bibinfo {author} {\bibfnamefont
  {T.~A.}\ \bibnamefont {Maier}}, \bibinfo {author} {\bibfnamefont {P.~C.}\
  \bibnamefont {Snijders}},\ and\ \bibinfo {author} {\bibfnamefont {H.~H.}\
  \bibnamefont {Weitering}},\ }\bibfield  {title} {\bibinfo {title}
  {Realization of a hole-doped mott insulator on a triangular silicon
  lattice},\ }\href {https://link.aps.org/doi/10.1103/PhysRevLett.119.266802}
  {\bibfield  {journal} {\bibinfo  {journal} {Phys. Rev. Lett.}\ }\textbf
  {\bibinfo {volume} {119}},\ \bibinfo {pages} {266802} (\bibinfo {year}
  {2017})}\BibitemShut {NoStop}%
\bibitem [{\citenamefont {Li}\ \emph {et~al.}(2013)\citenamefont {Li},
  \citenamefont {Höpfner}, \citenamefont {Schäfer}, \citenamefont
  {Blumenstein}, \citenamefont {Meyer}, \citenamefont {Bostwick}, \citenamefont
  {Rotenberg}, \citenamefont {Claessen},\ and\ \citenamefont
  {Hanke}}]{30limagnetic2013}%
  \BibitemOpen
  \bibfield  {author} {\bibinfo {author} {\bibfnamefont {G.}~\bibnamefont
  {Li}}, \bibinfo {author} {\bibfnamefont {P.}~\bibnamefont {Höpfner}},
  \bibinfo {author} {\bibfnamefont {J.}~\bibnamefont {Schäfer}}, \bibinfo
  {author} {\bibfnamefont {C.}~\bibnamefont {Blumenstein}}, \bibinfo {author}
  {\bibfnamefont {S.}~\bibnamefont {Meyer}}, \bibinfo {author} {\bibfnamefont
  {A.}~\bibnamefont {Bostwick}}, \bibinfo {author} {\bibfnamefont
  {E.}~\bibnamefont {Rotenberg}}, \bibinfo {author} {\bibfnamefont
  {R.}~\bibnamefont {Claessen}},\ and\ \bibinfo {author} {\bibfnamefont
  {W.}~\bibnamefont {Hanke}},\ }\bibfield  {title} {\bibinfo {title} {Magnetic
  order in a frustrated two-dimensional atom lattice at a semiconductor
  surface},\ }\href {https://doi.org/10.1038/ncomms2617} {\bibfield  {journal}
  {\bibinfo  {journal} {Nature Communications}\ }\textbf {\bibinfo {volume}
  {4}},\ \bibinfo {pages} {1620} (\bibinfo {year} {2013})}\BibitemShut
  {NoStop}%
\bibitem [{\citenamefont {Wu}\ \emph {et~al.}(2020)\citenamefont {Wu},
  \citenamefont {Ming}, \citenamefont {Smith}, \citenamefont {Liu},
  \citenamefont {Ye}, \citenamefont {Wang}, \citenamefont {Johnston},\ and\
  \citenamefont {Weitering}}]{26WuXuefengSuperconductivity2020}%
  \BibitemOpen
  \bibfield  {author} {\bibinfo {author} {\bibfnamefont {X.}~\bibnamefont
  {Wu}}, \bibinfo {author} {\bibfnamefont {F.}~\bibnamefont {Ming}}, \bibinfo
  {author} {\bibfnamefont {T.~S.}\ \bibnamefont {Smith}}, \bibinfo {author}
  {\bibfnamefont {G.}~\bibnamefont {Liu}}, \bibinfo {author} {\bibfnamefont
  {F.}~\bibnamefont {Ye}}, \bibinfo {author} {\bibfnamefont {K.}~\bibnamefont
  {Wang}}, \bibinfo {author} {\bibfnamefont {S.}~\bibnamefont {Johnston}},\
  and\ \bibinfo {author} {\bibfnamefont {H.~H.}\ \bibnamefont {Weitering}},\
  }\bibfield  {title} {\bibinfo {title} {Superconductivity in a hole-doped
  mott-insulating triangular adatom layer on a silicon surface},\ }\href
  {https://link.aps.org/doi/10.1103/PhysRevLett.125.117001} {\bibfield
  {journal} {\bibinfo  {journal} {Phys. Rev. Lett.}\ }\textbf {\bibinfo
  {volume} {125}},\ \bibinfo {pages} {117001} (\bibinfo {year}
  {2020})}\BibitemShut {NoStop}%
\bibitem [{\citenamefont {Ming}\ \emph {et~al.}(2023)\citenamefont {Ming},
  \citenamefont {Wu}, \citenamefont {Chen}, \citenamefont {Wang}, \citenamefont
  {Mai}, \citenamefont {Maier}, \citenamefont {Strockoz}, \citenamefont
  {Venderbos}, \citenamefont {González}, \citenamefont {Ortega}, \citenamefont
  {Johnston},\ and\ \citenamefont {Weitering}}]{27mingevidence2023}%
  \BibitemOpen
  \bibfield  {author} {\bibinfo {author} {\bibfnamefont {F.}~\bibnamefont
  {Ming}}, \bibinfo {author} {\bibfnamefont {X.}~\bibnamefont {Wu}}, \bibinfo
  {author} {\bibfnamefont {C.}~\bibnamefont {Chen}}, \bibinfo {author}
  {\bibfnamefont {K.~D.}\ \bibnamefont {Wang}}, \bibinfo {author}
  {\bibfnamefont {P.}~\bibnamefont {Mai}}, \bibinfo {author} {\bibfnamefont
  {T.~A.}\ \bibnamefont {Maier}}, \bibinfo {author} {\bibfnamefont
  {J.}~\bibnamefont {Strockoz}}, \bibinfo {author} {\bibfnamefont {J.~W.~F.}\
  \bibnamefont {Venderbos}}, \bibinfo {author} {\bibfnamefont {C.}~\bibnamefont
  {González}}, \bibinfo {author} {\bibfnamefont {J.}~\bibnamefont {Ortega}},
  \bibinfo {author} {\bibfnamefont {S.}~\bibnamefont {Johnston}},\ and\
  \bibinfo {author} {\bibfnamefont {H.~H.}\ \bibnamefont {Weitering}},\
  }\bibfield  {title} {\bibinfo {title} {Evidence for chiral superconductivity
  on a silicon surface},\ }\href {https://doi.org/10.1038/s41567-022-01889-1}
  {\bibfield  {journal} {\bibinfo  {journal} {Nature Physics}\ }\textbf
  {\bibinfo {volume} {19}},\ \bibinfo {pages} {500} (\bibinfo {year}
  {2023})}\BibitemShut {NoStop}%
\bibitem [{\citenamefont {Profeta}\ and\ \citenamefont
  {Tosatti}(2007)}]{28ProfetaTriangular2007}%
  \BibitemOpen
  \bibfield  {author} {\bibinfo {author} {\bibfnamefont {G.}~\bibnamefont
  {Profeta}}\ and\ \bibinfo {author} {\bibfnamefont {E.}~\bibnamefont
  {Tosatti}},\ }\bibfield  {title} {\bibinfo {title} {Triangular mott-hubbard
  insulator phases of sn/si(111)$ and sn/ge(111)$ surfaces},\ }\href
  {https://link.aps.org/doi/10.1103/PhysRevLett.98.086401} {\bibfield
  {journal} {\bibinfo  {journal} {Phys. Rev. Lett.}\ }\textbf {\bibinfo
  {volume} {98}},\ \bibinfo {pages} {086401} (\bibinfo {year}
  {2007})}\BibitemShut {NoStop}%
\bibitem [{\citenamefont {Modesti}\ \emph {et~al.}(2007)\citenamefont
  {Modesti}, \citenamefont {Petaccia}, \citenamefont {Ceballos}, \citenamefont
  {Vobornik}, \citenamefont {Panaccione}, \citenamefont {Rossi}, \citenamefont
  {Ottaviano}, \citenamefont {Larciprete}, \citenamefont {Lizzit},\ and\
  \citenamefont {Goldoni}}]{29ModestiInsulating2007}%
  \BibitemOpen
  \bibfield  {author} {\bibinfo {author} {\bibfnamefont {S.}~\bibnamefont
  {Modesti}}, \bibinfo {author} {\bibfnamefont {L.}~\bibnamefont {Petaccia}},
  \bibinfo {author} {\bibfnamefont {G.}~\bibnamefont {Ceballos}}, \bibinfo
  {author} {\bibfnamefont {I.}~\bibnamefont {Vobornik}}, \bibinfo {author}
  {\bibfnamefont {G.}~\bibnamefont {Panaccione}}, \bibinfo {author}
  {\bibfnamefont {G.}~\bibnamefont {Rossi}}, \bibinfo {author} {\bibfnamefont
  {L.}~\bibnamefont {Ottaviano}}, \bibinfo {author} {\bibfnamefont
  {R.}~\bibnamefont {Larciprete}}, \bibinfo {author} {\bibfnamefont
  {S.}~\bibnamefont {Lizzit}},\ and\ \bibinfo {author} {\bibfnamefont
  {A.}~\bibnamefont {Goldoni}},\ }\bibfield  {title} {\bibinfo {title}
  {Insulating ground state of
  $\mathrm{Sn}/\mathrm{Si}(111)\mathrm{\text{\ensuremath{-}}}(\sqrt{3}\ifmmode\times\else\texttimes\fi{}\sqrt{3})r30\ifmmode^\circ\else\textdegree\fi{}$},\
  }\href {https://link.aps.org/doi/10.1103/PhysRevLett.98.126401} {\bibfield
  {journal} {\bibinfo  {journal} {Phys. Rev. Lett.}\ }\textbf {\bibinfo
  {volume} {98}},\ \bibinfo {pages} {126401} (\bibinfo {year}
  {2007})}\BibitemShut {NoStop}%
\bibitem [{\citenamefont {J\"ager}\ \emph {et~al.}(2018)\citenamefont
  {J\"ager}, \citenamefont {Brand}, \citenamefont {Weber}, \citenamefont
  {Fanciulli}, \citenamefont {Dil}, \citenamefont {Pfn\"ur},\ and\
  \citenamefont {Tegenkamp}}]{32JageralphaSn2018}%
  \BibitemOpen
  \bibfield  {author} {\bibinfo {author} {\bibfnamefont {M.}~\bibnamefont
  {J\"ager}}, \bibinfo {author} {\bibfnamefont {C.}~\bibnamefont {Brand}},
  \bibinfo {author} {\bibfnamefont {A.~P.}\ \bibnamefont {Weber}}, \bibinfo
  {author} {\bibfnamefont {M.}~\bibnamefont {Fanciulli}}, \bibinfo {author}
  {\bibfnamefont {J.~H.}\ \bibnamefont {Dil}}, \bibinfo {author} {\bibfnamefont
  {H.}~\bibnamefont {Pfn\"ur}},\ and\ \bibinfo {author} {\bibfnamefont
  {C.}~\bibnamefont {Tegenkamp}},\ }\bibfield  {title} {\bibinfo {title}
  {$\ensuremath{\alpha}$-sn phase on si(111): Spin texture of a two-dimensional
  mott state},\ }\href {https://link.aps.org/doi/10.1103/PhysRevB.98.165422}
  {\bibfield  {journal} {\bibinfo  {journal} {Phys. Rev. B}\ }\textbf {\bibinfo
  {volume} {98}},\ \bibinfo {pages} {165422} (\bibinfo {year}
  {2018})}\BibitemShut {NoStop}%
\bibitem [{\citenamefont {Scalapino}(2012)}]{33ScalapinoAcommon2012}%
  \BibitemOpen
  \bibfield  {author} {\bibinfo {author} {\bibfnamefont {D.~J.}\ \bibnamefont
  {Scalapino}},\ }\bibfield  {title} {\bibinfo {title} {A common thread: The
  pairing interaction for unconventional superconductors},\ }\href
  {https://link.aps.org/doi/10.1103/RevModPhys.84.1383} {\bibfield  {journal}
  {\bibinfo  {journal} {Rev. Mod. Phys.}\ }\textbf {\bibinfo {volume} {84}},\
  \bibinfo {pages} {1383} (\bibinfo {year} {2012})}\BibitemShut {NoStop}%
\bibitem [{\citenamefont {Sigrist}\ and\ \citenamefont
  {Ueda}(1991)}]{34SigristPhenomenological1991}%
  \BibitemOpen
  \bibfield  {author} {\bibinfo {author} {\bibfnamefont {M.}~\bibnamefont
  {Sigrist}}\ and\ \bibinfo {author} {\bibfnamefont {K.}~\bibnamefont {Ueda}},\
  }\bibfield  {title} {\bibinfo {title} {Phenomenological theory of
  unconventional superconductivity},\ }\href
  {https://link.aps.org/doi/10.1103/RevModPhys.63.239} {\bibfield  {journal}
  {\bibinfo  {journal} {Rev. Mod. Phys.}\ }\textbf {\bibinfo {volume} {63}},\
  \bibinfo {pages} {239} (\bibinfo {year} {1991})}\BibitemShut {NoStop}%
\bibitem [{\citenamefont {Kallin}\ and\ \citenamefont
  {Berlinsky}(2016)}]{35kallinchiral2016}%
  \BibitemOpen
  \bibfield  {author} {\bibinfo {author} {\bibfnamefont {C.}~\bibnamefont
  {Kallin}}\ and\ \bibinfo {author} {\bibfnamefont {J.}~\bibnamefont
  {Berlinsky}},\ }\bibfield  {title} {\bibinfo {title} {Chiral
  superconductors},\ }\href {https://dx.doi.org/10.1088/0034-4885/79/5/054502}
  {\bibfield  {journal} {\bibinfo  {journal} {Reports on Progress in Physics}\
  }\textbf {\bibinfo {volume} {79}},\ \bibinfo {pages} {054502} (\bibinfo
  {year} {2016})}\BibitemShut {NoStop}%
\bibitem [{\citenamefont {Zampronio}\ and\ \citenamefont
  {Macr{\`{i}}}(2023)}]{36Zamproniochiral2023}%
  \BibitemOpen
  \bibfield  {author} {\bibinfo {author} {\bibfnamefont {V.}~\bibnamefont
  {Zampronio}}\ and\ \bibinfo {author} {\bibfnamefont {T.}~\bibnamefont
  {Macr{\`{i}}}},\ }\bibfield  {title} {\bibinfo {title} {Chiral
  superconductivity in the doped triangular-lattice {F}ermi-{H}ubbard model in
  two dimensions},\ }\href {https://doi.org/10.22331/q-2023-07-20-1061}
  {\bibfield  {journal} {\bibinfo  {journal} {{Quantum}}\ }\textbf {\bibinfo
  {volume} {7}},\ \bibinfo {pages} {1061} (\bibinfo {year} {2023})}\BibitemShut
  {NoStop}%
\bibitem [{\citenamefont {Cao}\ \emph {et~al.}(2018)\citenamefont {Cao},
  \citenamefont {Ayral}, \citenamefont {Zhong}, \citenamefont {Parcollet},
  \citenamefont {Manske},\ and\ \citenamefont
  {Hansmann}}]{37CaoXiaodongChiral2018}%
  \BibitemOpen
  \bibfield  {author} {\bibinfo {author} {\bibfnamefont {X.}~\bibnamefont
  {Cao}}, \bibinfo {author} {\bibfnamefont {T.}~\bibnamefont {Ayral}}, \bibinfo
  {author} {\bibfnamefont {Z.}~\bibnamefont {Zhong}}, \bibinfo {author}
  {\bibfnamefont {O.}~\bibnamefont {Parcollet}}, \bibinfo {author}
  {\bibfnamefont {D.}~\bibnamefont {Manske}},\ and\ \bibinfo {author}
  {\bibfnamefont {P.}~\bibnamefont {Hansmann}},\ }\bibfield  {title} {\bibinfo
  {title} {Chiral $d$-wave superconductivity in a triangular surface lattice
  mediated by long-range interaction},\ }\href
  {https://link.aps.org/doi/10.1103/PhysRevB.97.155145} {\bibfield  {journal}
  {\bibinfo  {journal} {Phys. Rev. B}\ }\textbf {\bibinfo {volume} {97}},\
  \bibinfo {pages} {155145} (\bibinfo {year} {2018})}\BibitemShut {NoStop}%
\bibitem [{\citenamefont {Ruby}\ \emph {et~al.}(2016)\citenamefont {Ruby},
  \citenamefont {Peng}, \citenamefont {von Oppen}, \citenamefont {Heinrich},\
  and\ \citenamefont {Franke}}]{38RubyOrbital2016}%
  \BibitemOpen
  \bibfield  {author} {\bibinfo {author} {\bibfnamefont {M.}~\bibnamefont
  {Ruby}}, \bibinfo {author} {\bibfnamefont {Y.}~\bibnamefont {Peng}}, \bibinfo
  {author} {\bibfnamefont {F.}~\bibnamefont {von Oppen}}, \bibinfo {author}
  {\bibfnamefont {B.~W.}\ \bibnamefont {Heinrich}},\ and\ \bibinfo {author}
  {\bibfnamefont {K.~J.}\ \bibnamefont {Franke}},\ }\bibfield  {title}
  {\bibinfo {title} {Orbital picture of yu-shiba-rusinov multiplets},\ }\href
  {https://link.aps.org/doi/10.1103/PhysRevLett.117.186801} {\bibfield
  {journal} {\bibinfo  {journal} {Phys. Rev. Lett.}\ }\textbf {\bibinfo
  {volume} {117}},\ \bibinfo {pages} {186801} (\bibinfo {year}
  {2016})}\BibitemShut {NoStop}%
\bibitem [{\citenamefont {Ménard}\ \emph {et~al.}(2015)\citenamefont
  {Ménard}, \citenamefont {Guissart}, \citenamefont {Brun}, \citenamefont
  {Pons}, \citenamefont {Stolyarov}, \citenamefont {Debontridder},
  \citenamefont {Leclerc}, \citenamefont {Janod}, \citenamefont {Cario},
  \citenamefont {Roditchev}, \citenamefont {Simon},\ and\ \citenamefont
  {Cren}}]{39menardcoherent2015}%
  \BibitemOpen
  \bibfield  {author} {\bibinfo {author} {\bibfnamefont {G.~C.}\ \bibnamefont
  {Ménard}}, \bibinfo {author} {\bibfnamefont {S.}~\bibnamefont {Guissart}},
  \bibinfo {author} {\bibfnamefont {C.}~\bibnamefont {Brun}}, \bibinfo {author}
  {\bibfnamefont {S.}~\bibnamefont {Pons}}, \bibinfo {author} {\bibfnamefont
  {V.~S.}\ \bibnamefont {Stolyarov}}, \bibinfo {author} {\bibfnamefont
  {F.}~\bibnamefont {Debontridder}}, \bibinfo {author} {\bibfnamefont {M.~V.}\
  \bibnamefont {Leclerc}}, \bibinfo {author} {\bibfnamefont {E.}~\bibnamefont
  {Janod}}, \bibinfo {author} {\bibfnamefont {L.}~\bibnamefont {Cario}},
  \bibinfo {author} {\bibfnamefont {D.}~\bibnamefont {Roditchev}}, \bibinfo
  {author} {\bibfnamefont {P.}~\bibnamefont {Simon}},\ and\ \bibinfo {author}
  {\bibfnamefont {T.}~\bibnamefont {Cren}},\ }\bibfield  {title} {\bibinfo
  {title} {Coherent long-range magnetic bound states in a superconductor},\
  }\href {https://doi.org/10.1038/nphys3508} {\bibfield  {journal} {\bibinfo
  {journal} {Nature Physics}\ }\textbf {\bibinfo {volume} {11}},\ \bibinfo
  {pages} {1013} (\bibinfo {year} {2015})}\BibitemShut {NoStop}%
\bibitem [{\citenamefont {Kim}\ \emph {et~al.}(2020)\citenamefont {Kim},
  \citenamefont {Rózsa}, \citenamefont {Schreyer}, \citenamefont {Simon},\
  and\ \citenamefont {Wiesendanger}}]{40kimlong-range2020}%
  \BibitemOpen
  \bibfield  {author} {\bibinfo {author} {\bibfnamefont {H.}~\bibnamefont
  {Kim}}, \bibinfo {author} {\bibfnamefont {L.}~\bibnamefont {Rózsa}},
  \bibinfo {author} {\bibfnamefont {D.}~\bibnamefont {Schreyer}}, \bibinfo
  {author} {\bibfnamefont {E.}~\bibnamefont {Simon}},\ and\ \bibinfo {author}
  {\bibfnamefont {R.}~\bibnamefont {Wiesendanger}},\ }\bibfield  {title}
  {\bibinfo {title} {Long-range focusing of magnetic bound states in
  superconducting lanthanum},\ }\href
  {https://doi.org/10.1038/s41467-020-18406-8} {\bibfield  {journal} {\bibinfo
  {journal} {Nature Communications}\ }\textbf {\bibinfo {volume} {11}},\
  \bibinfo {pages} {4573} (\bibinfo {year} {2020})}\BibitemShut {NoStop}%
\bibitem [{\citenamefont {Wang}\ and\ \citenamefont
  {Wang}(2004)}]{41WangImpurity2004}%
  \BibitemOpen
  \bibfield  {author} {\bibinfo {author} {\bibfnamefont {Q.-H.}\ \bibnamefont
  {Wang}}\ and\ \bibinfo {author} {\bibfnamefont {Z.~D.}\ \bibnamefont
  {Wang}},\ }\bibfield  {title} {\bibinfo {title} {Impurity and interface bound
  states in ${d}_{{x}^{2}\ensuremath{-}{y}^{2}}{+id}_{\mathrm{xy}}$ and
  ${p}_{x}{+ip}_{y}$ superconductors},\ }\href
  {https://link.aps.org/doi/10.1103/PhysRevB.69.092502} {\bibfield  {journal}
  {\bibinfo  {journal} {Phys. Rev. B}\ }\textbf {\bibinfo {volume} {69}},\
  \bibinfo {pages} {092502} (\bibinfo {year} {2004})}\BibitemShut {NoStop}%
\bibitem [{\citenamefont {Mashkoori}\ \emph {et~al.}(2017)\citenamefont
  {Mashkoori}, \citenamefont {Björnson},\ and\ \citenamefont
  {Black-Schaffer}}]{42mashkooriimpurity2017}%
  \BibitemOpen
  \bibfield  {author} {\bibinfo {author} {\bibfnamefont {M.}~\bibnamefont
  {Mashkoori}}, \bibinfo {author} {\bibfnamefont {K.}~\bibnamefont
  {Björnson}},\ and\ \bibinfo {author} {\bibfnamefont {A.~M.}\ \bibnamefont
  {Black-Schaffer}},\ }\bibfield  {title} {\bibinfo {title} {Impurity bound
  states in fully gapped d-wave superconductors with subdominant order
  parameters},\ }\href {https://doi.org/10.1038/srep44107} {\bibfield
  {journal} {\bibinfo  {journal} {Scientific Reports}\ }\textbf {\bibinfo
  {volume} {7}},\ \bibinfo {pages} {44107} (\bibinfo {year}
  {2017})}\BibitemShut {NoStop}%
\bibitem [{\citenamefont {Marchetti}\ \emph {et~al.}(2025)\citenamefont
  {Marchetti}, \citenamefont {Bunney}, \citenamefont {Di~Sante},\ and\
  \citenamefont {Rachel}}]{43MarchettiElectronic2025}%
  \BibitemOpen
  \bibfield  {author} {\bibinfo {author} {\bibfnamefont {L.}~\bibnamefont
  {Marchetti}}, \bibinfo {author} {\bibfnamefont {M.}~\bibnamefont {Bunney}},
  \bibinfo {author} {\bibfnamefont {D.}~\bibnamefont {Di~Sante}},\ and\
  \bibinfo {author} {\bibfnamefont {S.}~\bibnamefont {Rachel}},\ }\bibfield
  {title} {\bibinfo {title} {Electronic structure, spin-orbit interaction, and
  electron-phonon coupling of triangular adatom lattices on semiconductor
  substrates},\ }\href {https://link.aps.org/doi/10.1103/PhysRevB.111.125115}
  {\bibfield  {journal} {\bibinfo  {journal} {Phys. Rev. B}\ }\textbf {\bibinfo
  {volume} {111}},\ \bibinfo {pages} {125115} (\bibinfo {year}
  {2025})}\BibitemShut {NoStop}%
\bibitem [{\citenamefont {Cai}\ and\ \citenamefont
  {Zhang}(2025)}]{44caiDeciphering2025}%
  \BibitemOpen
  \bibfield  {author} {\bibinfo {author} {\bibfnamefont {Y.}~\bibnamefont
  {Cai}}\ and\ \bibinfo {author} {\bibfnamefont {R.-X.}\ \bibnamefont
  {Zhang}},\ }\bibfield  {title} {\bibinfo {title} {Deciphering chiral
  superconductivity via impurity bound states},\ }\href
  {https://arxiv.org/abs/2506.20842} {\bibfield  {journal} {\bibinfo  {journal}
  {arXiv preprint arXiv:2506.20842}\ } (\bibinfo {year} {2025})}\BibitemShut
  {NoStop}%
\bibitem [{\citenamefont {Wolf}\ \emph {et~al.}(2022)\citenamefont {Wolf},
  \citenamefont {Di~Sante}, \citenamefont {Schwemmer}, \citenamefont
  {Thomale},\ and\ \citenamefont {Rachel}}]{45WolfTriplet2022}%
  \BibitemOpen
  \bibfield  {author} {\bibinfo {author} {\bibfnamefont {S.}~\bibnamefont
  {Wolf}}, \bibinfo {author} {\bibfnamefont {D.}~\bibnamefont {Di~Sante}},
  \bibinfo {author} {\bibfnamefont {T.}~\bibnamefont {Schwemmer}}, \bibinfo
  {author} {\bibfnamefont {R.}~\bibnamefont {Thomale}},\ and\ \bibinfo {author}
  {\bibfnamefont {S.}~\bibnamefont {Rachel}},\ }\bibfield  {title} {\bibinfo
  {title} {Triplet superconductivity from nonlocal coulomb repulsion in an
  atomic sn layer deposited onto a si(111) substrate},\ }\href
  {https://link.aps.org/doi/10.1103/PhysRevLett.128.167002} {\bibfield
  {journal} {\bibinfo  {journal} {Phys. Rev. Lett.}\ }\textbf {\bibinfo
  {volume} {128}},\ \bibinfo {pages} {167002} (\bibinfo {year}
  {2022})}\BibitemShut {NoStop}%
\bibitem [{\citenamefont {Biderang}\ \emph {et~al.}(2022)\citenamefont
  {Biderang}, \citenamefont {Zare},\ and\ \citenamefont
  {Sirker}}]{46BiderangTopological2022}%
  \BibitemOpen
  \bibfield  {author} {\bibinfo {author} {\bibfnamefont {M.}~\bibnamefont
  {Biderang}}, \bibinfo {author} {\bibfnamefont {M.-H.}\ \bibnamefont {Zare}},\
  and\ \bibinfo {author} {\bibfnamefont {J.}~\bibnamefont {Sirker}},\
  }\bibfield  {title} {\bibinfo {title} {Topological superconductivity in
  sn/si(111) driven by nonlocal coulomb interactions},\ }\href
  {https://link.aps.org/doi/10.1103/PhysRevB.106.054514} {\bibfield  {journal}
  {\bibinfo  {journal} {Phys. Rev. B}\ }\textbf {\bibinfo {volume} {106}},\
  \bibinfo {pages} {054514} (\bibinfo {year} {2022})}\BibitemShut {NoStop}%
\bibitem [{\citenamefont {Kim}\ and\ \citenamefont
  {Pereg-Barnea}(2023)}]{47KimInterplay2023}%
  \BibitemOpen
  \bibfield  {author} {\bibinfo {author} {\bibfnamefont {K.~W.}\ \bibnamefont
  {Kim}}\ and\ \bibinfo {author} {\bibfnamefont {T.}~\bibnamefont
  {Pereg-Barnea}},\ }\bibfield  {title} {\bibinfo {title} {Interplay between
  superconductivity and magnetism in triangular lattices},\ }\href
  {https://link.aps.org/doi/10.1103/PhysRevB.108.195113} {\bibfield  {journal}
  {\bibinfo  {journal} {Phys. Rev. B}\ }\textbf {\bibinfo {volume} {108}},\
  \bibinfo {pages} {195113} (\bibinfo {year} {2023})}\BibitemShut {NoStop}%
\bibitem [{\citenamefont {Pientka}\ \emph {et~al.}(2013)\citenamefont
  {Pientka}, \citenamefont {Glazman},\ and\ \citenamefont {von
  Oppen}}]{48PientkaTopological2013}%
  \BibitemOpen
  \bibfield  {author} {\bibinfo {author} {\bibfnamefont {F.}~\bibnamefont
  {Pientka}}, \bibinfo {author} {\bibfnamefont {L.~I.}\ \bibnamefont
  {Glazman}},\ and\ \bibinfo {author} {\bibfnamefont {F.}~\bibnamefont {von
  Oppen}},\ }\bibfield  {title} {\bibinfo {title} {Topological superconducting
  phase in helical shiba chains},\ }\href
  {https://link.aps.org/doi/10.1103/PhysRevB.88.155420} {\bibfield  {journal}
  {\bibinfo  {journal} {Phys. Rev. B}\ }\textbf {\bibinfo {volume} {88}},\
  \bibinfo {pages} {155420} (\bibinfo {year} {2013})}\BibitemShut {NoStop}%
\bibitem [{\citenamefont {Kaladzhyan}\ \emph {et~al.}(2016)\citenamefont
  {Kaladzhyan}, \citenamefont {Bena},\ and\ \citenamefont
  {Simon}}]{49kaladzhyanasymptotic2016}%
  \BibitemOpen
  \bibfield  {author} {\bibinfo {author} {\bibfnamefont {V.}~\bibnamefont
  {Kaladzhyan}}, \bibinfo {author} {\bibfnamefont {C.}~\bibnamefont {Bena}},\
  and\ \bibinfo {author} {\bibfnamefont {P.}~\bibnamefont {Simon}},\ }\bibfield
   {title} {\bibinfo {title} {Asymptotic behavior of impurity-induced bound
  states in low-dimensional topological superconductors},\ }\href
  {https://dx.doi.org/10.1088/0953-8984/28/48/485701} {\bibfield  {journal}
  {\bibinfo  {journal} {Journal of Physics: Condensed Matter}\ }\textbf
  {\bibinfo {volume} {28}},\ \bibinfo {pages} {485701} (\bibinfo {year}
  {2016})}\BibitemShut {NoStop}%
\bibitem [{\citenamefont {Jemander}\ \emph {et~al.}(2001)\citenamefont
  {Jemander}, \citenamefont {Lin}, \citenamefont {Zhang}, \citenamefont
  {Uhrberg},\ and\ \citenamefont {Hansson}}]{50jemanderstm2001}%
  \BibitemOpen
  \bibfield  {author} {\bibinfo {author} {\bibfnamefont {S.~T.}\ \bibnamefont
  {Jemander}}, \bibinfo {author} {\bibfnamefont {N.}~\bibnamefont {Lin}},
  \bibinfo {author} {\bibfnamefont {H.~M.}\ \bibnamefont {Zhang}}, \bibinfo
  {author} {\bibfnamefont {R.~I.~G.}\ \bibnamefont {Uhrberg}},\ and\ \bibinfo
  {author} {\bibfnamefont {G.~V.}\ \bibnamefont {Hansson}},\ }\bibfield
  {title} {\bibinfo {title} {An {STM} study of the surface defects of the
  (3×3)-sn/si(111) surface},\ }\href
  {https://www.sciencedirect.com/science/article/pii/S0039602800011006}
  {\bibfield  {journal} {\bibinfo  {journal} {Surface Science}\ }\textbf
  {\bibinfo {volume} {475}},\ \bibinfo {pages} {181} (\bibinfo {year}
  {2001})}\BibitemShut {NoStop}%
\end{thebibliography}%

\end{document}